\documentclass[twocolumn,twocolappendix]{aastex631}

\usepackage{graphicx}	
\usepackage{amsmath}	
\usepackage{subfigure}
\usepackage{natbib}
\usepackage{color}
\usepackage{tabularx}
\usepackage{longtable}
\usepackage{bm}
\usepackage{threeparttable}
\usepackage{enumerate}
\usepackage[normalem]{ulem}
\usepackage{verbatim}
\usepackage{cleveref}

\newcommand{\code}[1]{{\texttt{#1}}}
\newcommand{\tess}{{\it TESS}}

\newcommand{\gaia}{{\it Gaia}}

\newcommand{\tar}{TOI-5575}

\begin{document}

\title{A New Brown Dwarf Orbiting an M star and An Investigation on the Eccentricity Distribution of Transiting Long-Period Brown Dwarfs}

\correspondingauthor{Tianjun Gan}
\email{tianjungan@gmail.com}

\author[0000-0002-4503-9705]{Tianjun~Gan}
\affil{Department of Astronomy, Westlake University, Hangzhou 310030, Zhejiang Province, China}

\author[0000-0001-9291-5555]{Charles Cadieux}
\affiliation{Universit\'e de Montr\'eal, D\'epartement de Physique, IREX, Montr\'eal, QC H3C 3J7, Canada}

\author[0000-0001-9564-6186]{Shigeru Ida}
\affil{Earth-Life Science Institute, Institute of Science Tokyo, Meguro, Tokyo 152-8550, Japan}

\author[0000-0002-6937-9034]{Sharon X. Wang} 
\affil{Department of Astronomy, Tsinghua University, Beijing 100084, China}

\author[0000-0001-8317-2788]{Shude Mao} 
\affil{Department of Astronomy, Westlake University, Hangzhou 310030, Zhejiang Province, China}

\author[0000-0001-5695-8734]{Zitao Lin} 
\affil{Department of Astronomy, Tsinghua University, Beijing 100084, China}

\author[0000-0002-3481-9052]{Keivan G. Stassun} 
\affil{Department of Physics and Astronomy, Vanderbilt University, 6301 Stevenson Center Ln., Nashville, TN 37235, USA}
\affil{Department of Physics, Fisk University, 1000 17th Avenue North, Nashville, TN 37208, USA}

\author[0000-0002-6523-9536]{Adam J.~Burgasser} 
\affil{Department of Astronomy \& Astrophysics, University of California, San Diego, 9500 Gilman Dr, La Jolla, CA 92093, USA}

\author[0000-0002-2532-2853]{Steve~B.~Howell} 
\affiliation{NASA Ames Research Center, Moffett Field, CA 94035, USA}

\author[0000-0002-2361-5812]{Catherine A. Clark} 
\affil{NASA Exoplanet Science Institute-Caltech/IPAC, Pasadena, CA 91125, USA}

\author[0000-0003-0647-6133]{Ivan A. Strakhov} 
\affil{Sternberg Astronomical Institute, Lomonosov Moscow State University, 119992 Universitetskii prospekt 13, Moscow, Russia}

\author[0000-0001-6981-8722]{Paul Benni} 
\affil{Acton Sky Portal (private observatory), Acton, MA USA}

\author[0000-0003-2058-6662]{George~R.~Ricker}
\affil{Department of Physics and Kavli Institute for Astrophysics and Space Research, Massachusetts Institute of Technology, Cambridge, MA 02139, USA}

\author[0000-0001-6763-6562]{Roland~Vanderspek}
\affil{Department of Physics and Kavli Institute for Astrophysics and Space Research, Massachusetts Institute of Technology, Cambridge, MA 02139, USA}

\author[0000-0001-9911-7388]{David~W.~Latham}
\affil{Center for Astrophysics ${\rm \mid}$ Harvard {\rm \&} Smithsonian, 60 Garden Street, Cambridge, MA 02138, USA}

\author[0000-0002-6892-6948]{Sara~Seager}
\affil{Department of Physics and Kavli Institute for Astrophysics and Space Research, Massachusetts Institute of Technology, Cambridge, MA 02139, USA}
\affil{Department of Earth, Atmospheric and Planetary Science, Massachusetts Institute of Technology, 77 Massachusetts Avenue, Cambridge, MA 02139, USA}
\affil{Department of Aeronautics and Astronautics, MIT, 77 Massachusetts Avenue, Cambridge, MA 02139, USA}

\author[0000-0002-4265-047X]{Joshua~N.~Winn}
\affil{Department of Astrophysical Sciences, Princeton University, 4 Ivy Lane, Princeton, NJ 08544, USA}

\author[0000-0002-4715-9460]{Jon~M.~Jenkins}
\affil{NASA Ames Research Center, Moffett Field, CA 94035, USA}

\author[0000-0002-0111-1234]{Luc Arnold}
\affiliation{Canada–France–Hawaii Telescope, 65-1238 Mamalahoa Hwy, Kamuela, HI 96743, USA}

\author[0000-0003-3506-5667]{\'Etienne Artigau}
\affiliation{Universit\'e de Montr\'eal, D\'epartement de Physique, IREX, Montr\'eal, QC H3C 3J7, Canada}
\affiliation{Observatoire du Mont-M\'egantic, Universit\'e de Montr\'eal, Montr\'eal, QC H3C 3J7, Canada}

\author[0000-0002-9003-484X]{David~Charbonneau} 
\affil{Center for Astrophysics ${\rm \mid}$ Harvard {\rm \&} Smithsonian, 60 Garden Street, Cambridge, MA 02138, USA}

\author[0000-0001-6588-9574]{Karen A. Collins} 
\affil{Center for Astrophysics ${\rm \mid}$ Harvard {\rm \&} Smithsonian, 60 Garden Street, Cambridge, MA 02138, USA}

\author[0000-0003-4166-4121]{Neil J. Cook}
\affiliation{Universit\'e de Montr\'eal, D\'epartement de Physique, IREX, Montr\'eal, QC H3C 3J7, Canada}

\author[0000-0002-7564-6047]{Zo{\" e} L. de Beurs} 
\affil{Department of Earth, Atmospheric and Planetary Science, Massachusetts Institute of Technology, 77 Massachusetts Avenue, Cambridge, MA 02139, USA}
\affil{NSF Graduate Research Fellow, MIT Presidential Fellow, MIT Collamore-Rogers Fellow, MIT Teaching Development Fellow}

\author[0009-0002-9833-0667]{Sarah~J.~Deveny} 
\affiliation{Bay Area Environmental Research Institute, Moffett Field, CA 94035, USA}
\affiliation{NASA Ames Research Center, Moffett Field, CA 94035, USA}

\author{John~P.~Doty} 
\affiliation{Noqsi Aerospace Ltd., 15 Blanchard Avenue, Billerica, MA 01821, USA}

\author[0000-0001-5485-4675]{Ren\'e Doyon}
\affiliation{Universit\'e de Montr\'eal, D\'epartement de Physique, IREX, Montr\'eal, QC H3C 3J7, Canada}
\affiliation{Observatoire du Mont-M\'egantic, Universit\'e de Montr\'eal, Montr\'eal, QC H3C 3J7, Canada}

\author[0000-0001-7746-5795]{Colin Littlefield} 
\affiliation{Bay Area Environmental Research Institute, Moffett Field, CA 94035, USA}
\affiliation{NASA Ames Research Center, Moffett Field, CA 94035, USA}

\author[0000-0001-9227-8349]{Tyler Pritchard} 
\affiliation{NASA Goddard Space Flight Center, 8800 Greenbelt Road, Greenbelt, MD 20771, USA}

\author[0009-0006-7023-1199]{Gabrielle Ross} 
\affiliation{Department of Astrophysical Sciences, Princeton University, 4 Ivy Lane, Princeton, NJ 08544, USA}

\author[0000-0002-1836-3120]{Avi Shporer} 
\affil{Department of Physics and Kavli Institute for Astrophysics and Space Research, Massachusetts Institute of Technology, Cambridge, MA 02139, USA}

\author[0000-0002-9807-5435]{Christopher R.~Theissen} 
\affil{Department of Astronomy \& Astrophysics, University of California, San Diego, 9500 Gilman Dr, La Jolla, CA 92093, USA}

\author[0000-0003-2053-0749]{Benjamin M.~Tofflemire}
\affil{SETI Institute, Mountain View, CA 94043 USA/NASA Ames Research Center, Moffett Field, CA 94035 USA}

\author[0000-0001-7246-5438]{Andrew Vanderburg} 
\affil{Department of Physics and Kavli Institute for Astrophysics and Space Research, Massachusetts Institute of Technology, Cambridge, MA 02139, USA}
\affil{Sloan Fellow}

\author[0000-0002-3555-8464]{David Watanabe} 
\affiliation{Planetary Discoveries, Valencia, CA 91354, USA}




\begin{abstract}

The orbital eccentricities of brown dwarfs encode valuable information of their formation and evolution history, providing insights into whether they resemble giant planets or stellar binaries. Here, we report the discovery of \tar b, a long-period, massive brown dwarf orbiting a low-mass M5V star ($\rm 0.21\pm0.02\,M_\odot$) delivered by the TESS mission. The companion has a mass and radius of $\rm 72.4\pm4.1\,M_J$ and $\rm 0.84\pm0.07\,R_J$ on a 32-day moderately eccentric orbit ($e=0.187\pm0.002$), making it the third highest-mass-ratio transiting brown dwarf system known to date. Building on this discovery, we investigate the eccentricity distributions of a sample of transiting long-period ($10\leq P\lesssim 1000$\,days, $\sim$0.1-1.5\,AU) giant planets, brown dwarfs and low-mass stars. We find that brown dwarfs exhibit an eccentricity behavior nearly identical to that of giant planets: a preference for circular orbits with a long tail toward high eccentricities. Such a trend contrasts sharply with direct imaging findings, where cold (5-100\,AU) brown dwarfs and giant planets display distinct eccentricity distributions. Our results suggest that transiting long-period brown dwarfs and giant planets probably 1) form in different routes at exterior orbits but undergo analogous dynamical evolution processes and migrate inwards; or 2) both contain two sub-groups, one with widely spread eccentricities while the other has circular orbits, that jointly sculpt the eccentricity distributions. The low-mass-star systems appear to be a distinctive population, showing a peak eccentricity at about 0.3, akin to more massive stellar binaries.

\end{abstract}



\section{Introduction} \label{sec:intro}

As intermediate objects between giant planets and low-mass stars, brown dwarfs (BDs) have masses above the deuterium burning limit \citep[$\sim 13$\,M$_{\rm J}$,][]{Burgasser2003,Spiegel2011} and below the threshold of hydrogen fusion \citep[$\sim 80$\,M$_{\rm J}$,][]{Laughlin1997}. Early blind radial velocity (RV) surveys have identified that brown dwarfs around Sun-like stars within 3\,AU are much rarer compared to planets and stellar companions, manifesting as a ``brown dwarf desert'' \citep{Marcy2000}. Such an occurrence rate deficit suggests that brown dwarfs are probably more challenging to form compared with the other two categories or that subsequent dynamical processes are unlikely to push brown dwarfs inwards after formation \citep{Grether2006}. Nevertheless, the number of transiting brown dwarfs has been enlarging dramatically during the past decade \citep[e.g.,][]{Artigau2021,Psaridi2022,Carmichael2022,Lin2023,Henderson2024,Larsen2025}. It remains under debate how such massive objects, in particular those having orbital periods $P\geq 10$~days, ended up at their current locations. 

Regarding the origins of warm Jupiters, three hypotheses have been postulated including in-situ formation, disk-driven migration, and high-eccentricity migration \citep{Dawson2018}. One route to probe the predominant scenario of the evolution channel is investigating the orbital eccentricities \citep[e.g.,][]{Dawson2012,Dong2021}, on which different mechanisms have different predictions. After initially forming through either core accretion \citep{Pollack1996} or gravitational instability \citep{Boss2000}, giant planets are expected to undergo dynamical interactions with other bodies or the disk \citep{Dawson2018}, and migrate inwards \citep{Lin1996}. During their evolution, giant planets could gather angular momentum and develop orbital eccentricities through, for example, planet-planet scattering \citep{Rasio1996,Dawson2013,Petrovich2016}, planet-disk interactions \citep{Goldreich2003} and Kozai–Lidov perturbations with outer massive companions \citep{Naoz2011}. Recent work from \cite{Alqasim2025} looked into transiting long-period giant planet systems with $P>10$~days, and identified a positive correlation between eccentricity and stellar metallicity (see also \citealt{Dawson2013}) but found no correlation between eccentricity and the presence of a second stellar companion (see also \citealt{Stevenson2025}), suggesting that planet-planet scattering \citep{Naoz2011,Beauge2012,Carrera2019} is more likely to be the way that produce warm Jupiters instead of the Kozai-Lidov effect \citep{Fabrycky2007}. On the other hand, stellar binaries are supposed to form through protostellar core or disk fragmentation \citep{Kratter2016} instead and migrate inwards driven by Kozai–Lidov interactions \citep{Eggleton2006,Naoz2014} or accreting from the circumbinary disc \citep{Tokovinin2020}. All these formation and dynamical processes are expected to conjunctly shape the final eccentricity distribution of the companion. 

However, brown dwarfs remain poorly understood on their evolution due to the limited number of detections that impede statistical studies, leaving a fundamental question unresolved: do brown dwarfs follow the routes of giant planets or more approximate to stellar binaries?  Early work from \cite{Ma2014} reported that brown dwarfs with masses above and below $42.5$\,M$_{\rm J}$ exhibit different eccentricity distributions: the massive population behaves like stellar binaries while the less massive group is more alike to planets, implying two different formation channels. \cite{Ma2014} made use of brown dwarfs mostly detected by radial velocity surveys and thus only have minimum masses available. According to recent results from Gaia astrometry, it turns out that several previously RV-detected brown dwarfs probably require reclassification as they are no longer located in the sub-stellar mass regime with true masses above $80$\,M$_{\rm J}$ \citep{Unger2023}. More recently, \cite{Vowell2025} questioned the existence of the mass transition point by looking into the mass ratio and metallicity distributions of brown dwarfs. Moving outwards, \cite{Bowler2020} found that directly-imaged brown dwarfs and giant planets between 5 and 100\,AU have significantly different eccentricity preferences (see also \citealt{Nagpal2023,Doo2023}). Cold Jupiters tend to exhibit small eccentricities peaking at around 0.05-0.25 while wide-orbit brown dwarfs have a broad eccentricity distribution, favoring higher values ($e\sim$~0.6-0.9). It is yet unclear if inner Jupiter-like planets and brown dwarfs follow the same trend. The increasing number of transiting brown dwarfs now enables a glimpse into their orbital eccentricity distribution and compare that with other two companion classes, which could in turn offer insights into their origins.

In this paper, we first report the discovery of \hbox{TOI-5575\,b}, a long-period massive brown dwarf around a low-mass star. Afterwards, we carry out population-level comparative studies between transiting long-period gas giants, brown dwarfs and low-mass stars. The paper is organized as follows. In Section~\ref{Observations}, we describe all observations we collected to confirm \hbox{TOI-5575\,b}. We perform the stellar characterization and a joint analysis of photometric and RV data in Section~\ref{Analysis}. Section~\ref{Eccentricity_Distribution} details how we select the samples, and conduct the eccentricity distribution comparison. We discuss our results in Section~\ref{Discussions} and conclude our findings in Section~\ref{Conclusions}.


\begin{table}[h]\scriptsize
    \caption{Summary of stellar properties for \tar.}
    \begin{tabular}{lll}
        \hline\hline
        Parameter       &Value       &Ref. \\\hline
        \it{Main identifiers}                    \\
         TIC                     &$160162137$   &$\rm TIC\ V8.2^{[1]}$\\
         \gaia\ ID            &$1724291831608394624$ &\gaia\ DR3$^{[2]}$\\
         \it{Equatorial Coordinates} \\
         $\alpha_{\rm J2015.5}$    &\ \ 16:04:59.65 &$\rm TIC\ V8.2$\\
         $\delta_{\rm J2015.5}$    &+85:12:17.01   &$\rm TIC\ V8.2$ \\
         \it{Photometric properties}\\
         $\tess$\ (mag)           &$14.013\pm0.007$   &$\rm TIC\ V8.2$  \\
         $V$ (mag) &$17.010\pm0.200$ &$\rm TIC\ V8.2$\\
         $G$ (mag)           &$15.352\pm0.001$   &\gaia\ DR3   \\
         $G_{\rm BP}$ (mag)           &$17.102\pm0.004$   &\gaia\ DR3   \\
         $G_{\rm RP}$ (mag)           &$14.093\pm0.001$   &\gaia\ DR3   \\
         $J$\ (mag)                    &$12.281\pm0.021$   &2MASS$^{[3]}$\\
         $H$\ (mag)                    &$11.723\pm0.017$   &2MASS \\
         $K$\ (mag)                    &$11.436\pm0.019$    &2MASS \\
         $W$1 (mag)                   &$11.266\pm0.023$   &WISE$^{[4]}$ \\
         $W$2 (mag)                   &$11.061\pm0.020$   &WISE \\
         $W$3 (mag)                   &$10.815\pm0.060$   &WISE \\
         \it{Astrometric properties}\\
         $\varpi$ (mas)              &$17.784\pm0.027$  &\gaia\ DR3  \\
         $\mu_{\alpha}\ ({\rm mas~yr^{-1}})$     &$-19.181\pm0.032$   &\gaia\ DR3   \\
         $\mu_{\delta}\ ({\rm mas~yr^{-1}})$     &$131.205\pm0.034$   &\gaia\ DR3  \\
         RV\ (km~s$^{-1}$)         &$-38.8\pm 4.2$ &\gaia\ DR3  \\
         \it{Stellar parameters} \\
         Spectral Type &$\rm M5.0\pm 0.5$ & This work\\
         Distance (pc)                &$56.2\pm 0.1$  &This work     \\
         $M_{\ast}\ ({\rm M_{\odot}})$ &$0.21\pm 0.02$ &This work       \\
         $R_{\ast}\ ({\rm R_{\odot}})$ &$0.24\pm 0.02$ &This work       \\
         $\log g_{\ast}\ ({\rm cgs})$       &$5.00\pm 0.04$  &This work        \\
         $T_{\rm eff}\ ({\rm K})$           &$3115\pm 100$  &This work       \\
         $\rm [Fe/H]$ (dex)  &$-0.21\pm 0.07$ &This work \\
         \hline\hline 
    \end{tabular}
    \begin{tablenotes}
    \item[1]  [1]\cite{Stassun2019tic}; [2]\cite{Gaia2023}; [3]\cite{Cutri2003}; [4]\cite{wright2010}.
    \end{tablenotes}
    \label{starparam}
\end{table}

\section{Observations}\label{Observations}

\subsection{TESS and Ground-Based Photometry}\label{photometry}

\tar\ (TIC 160162137) was first observed by the Transiting Exoplanets Survey Satellite \citep[TESS;][]{Ricker2015} in six Sectors every 30 minutes between 2019 and 2020. Subsequent monitoring during the First Extended Mission covered five additional sectors at a 10-minute cadence, followed by five more sectors at a 2-minute cadence during the ongoing Second Extended Mission. Due to the faintness of the host star ($T{\rm mag}=14.0$), we note that the photometry of \tar\ during the TESS Primary Mission is not available since the Quick Look Pipeline \citep[QLP;][]{QLP2020a,QLP2020b} produce light curves of objects down to $T{\rm mag}= 13.5$ and only those down to $T{\rm mag}= 10.5$ are searched for transits. The candidate was not alerted until the initiation of the QLP faint-star search program \citep{Kunimoto2022}.

To homogeneously extract the light curves from the TESS Primary Mission, we performed uniform aperture photometry using the \code{lightkurve} package \citep{lightkurvecollaboration}, following the procedures in \cite{Gan2023TOI4201}. In short, we first obtain the photometry using a custom $3\times3$ box aperture centering at the target star, and then subtract it with the time-series of the background. The light curves from the TESS First and Second Extended Missions were extracted using the QLP and by the Science Processing Operations Center \citep[SPOC;][]{Jenkins2016,Stumpe2012,Stumpe2014,Smith2012}, respectively. For the SPOC data, we use the simple aperture photometry without light dilution correction. We then employ the \code{celerite} algorithm \citep{Foreman2017} to detrend the light curve by fitting a Gaussian Process (GP) model with a Mat\'{e}rn-3/2 kernel to the out-of-transit data. The detrending was carried out independently for TESS data obtained during three Missions since the observational cadences and the sources of photometry are different. Figure~\ref{tesslc} presents the raw and detrended TESS light curves.  

In addition to the TESS data, we collected a ground-based light curve using the 0.36-m Acton-Sky-Portal (ASP) telescope on 2023 December 14. The observation was conducted in a blue-blocking clear (CBB) filter and fully covered the transit event. The ASP telescope is equipped with an SBIG Aluma CCD47-10 camera with a plate scale of 1 arcsec per pixel, providing a $17'\times17'$ field of view. We performed the photometric analysis using the \code{AstroImageJ} package \citep{Collins2017} with an uncontaminated $5''$ aperture. 


\subsection{Spectroscopic Follow-up Observations}

\subsubsection{CFHT/SPIRou Radial Velocities}

We obtained a total of 27 SPIRou \citep[SpectroPolarim\`etre InfraROUge;][]{Donati2020} spectra of \tar\ between UT 2023 May 1 and UT 2023 June 30 to determine the mass of the transiting companion. SPIRou is a near-infrared spectropolarimeter installed on the 3.6-m Canada-France-Hawaii Telescope (CFHT), offering a spectral resolution of $R\approx 75,000$ and covering the 0.95--2.35\,$\mu$m wavelength range. The observations were carried out over thirteen nights usually with two consecutive 1200s exposures per night under an environment condition of airmass around 2.4 and seeing about $0.6\arcsec$. The median signal-to-noise ratio (SNR) of our spectra is 16 per pixel in the center of the $H$ band ($\sim$1.7\,$\mu$m).

\begin{figure*}
\centering
\includegraphics[width=0.49\textwidth]{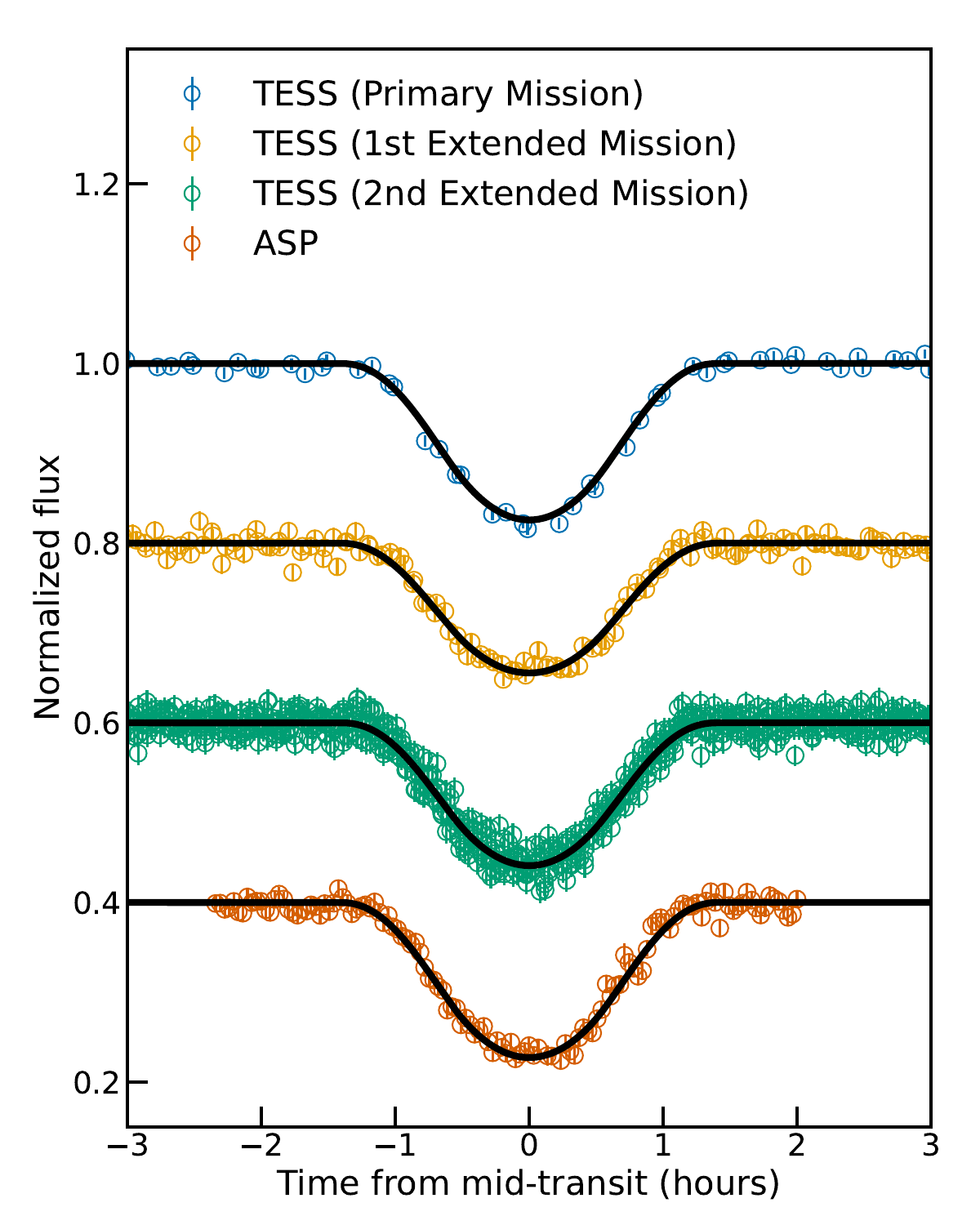}
\includegraphics[width=0.49\textwidth]{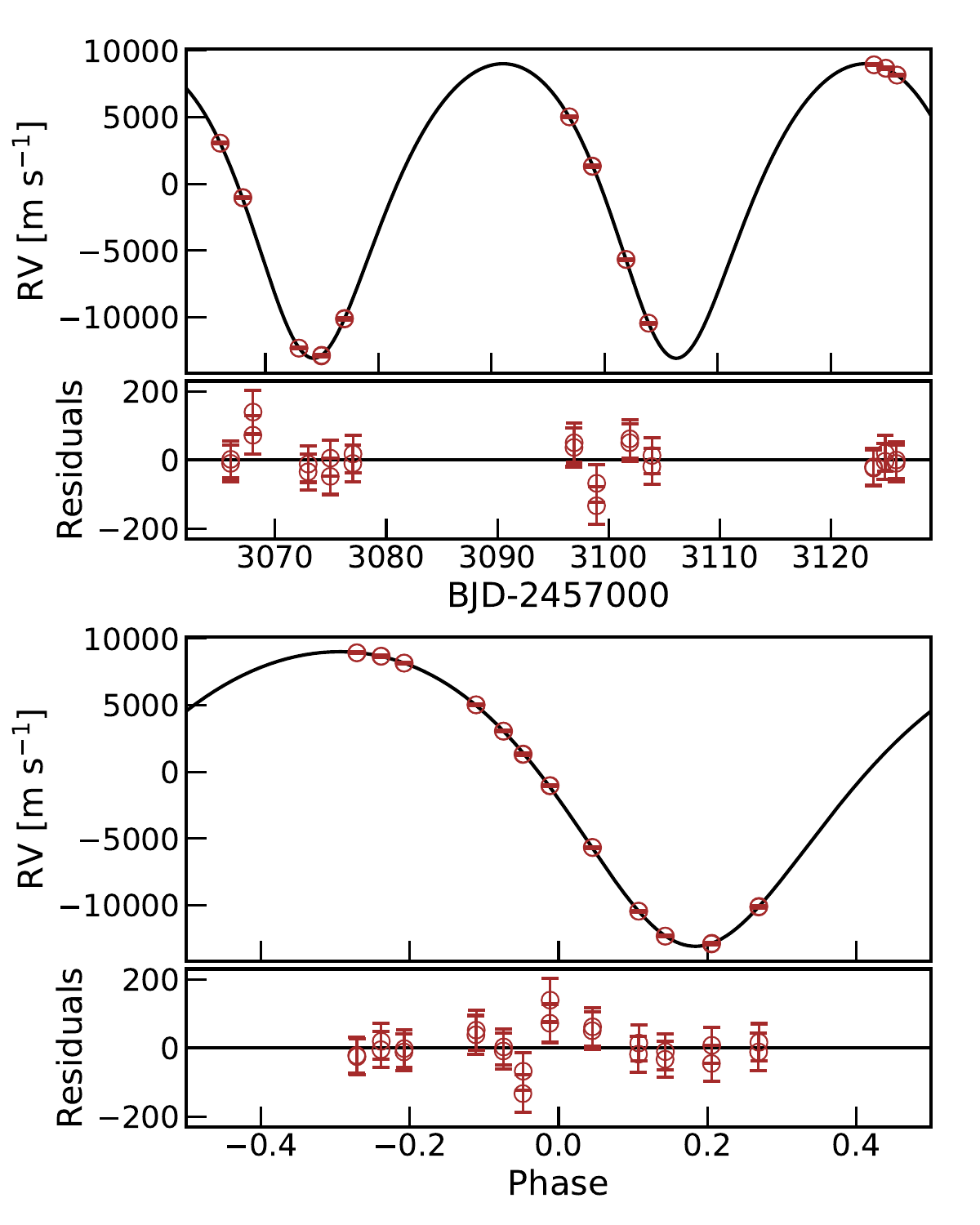}
\caption{{\it Left panel:} The TESS and ASP light curves folded in phase with the transit ephemeris from the joint analysis along with the best-fit transit models. {\it Right panel:} The full SPIRou RV timeseries and the phase-folded RVs. The error bars presenting here are the quadrature sum of the original RV uncertainties and the instrument jitter. The black solid line is the best-fit Keplerian model. The residuals are shown below each subplot.} 
\label{tess_joint_fit}
\end{figure*}

The raw SPIRou data were reduced with the \code{APERO} pipeline \citep{Cook2022}, which produced telluric-corrected spectra. We then measured the radial velocities from these telluric-corrected spectra through the line-by-line (LBL) method of \cite{Artigau2022}, employing a high SNR template of Gliese\,699 (M4V) observed with SPIRou instead of TOI-5575 to improve the precision. We discarded three epochs with outlying RV due to a low SNR below 5, leaving a final set of 24 measurements taken over 12 independent nights (see Figure~\ref{tess_joint_fit}). We measure a systemic velocity of about $-45.9$\,km\,s$^{-1}$ for \tar, which agrees with the result of $-38.8\pm4.2$\,km\,s$^{-1}$ from Gaia \citep{Gaia2023}. Table~\ref{RVtable} lists the final SPIRou RV measurements that reached a median precision of 18.5\,m\,s$^{-1}$.

\subsubsection{Keck/NIRES}

TOI-5575 was observed with the Near-Infrared Echellette Spectrometer (NIRES; \citealt{2004SPIE.5492.1295W}) mounted on the Keck~II 10m telescope on UT 2025 May 16 in clear conditions with 1$\arcsec$ seeing. NIRES provides 1.0--2.4~$\mu$m spectroscopy in four cross-dispersed orders at an average resolution of $\lambda/\Delta\lambda$ $\approx$ 2700 for its fixed 0$\farcs$55-wide slit (Figure~\ref{fig:sed}). The source was observed over two series of ABBA sequences (8 exposures) nodding along the 10$\arcsec$ slit, with 120~s exposures. Data were reduced using a modified version of Spextool \citep{2004PASP..116..362C} following standard extraction procedures and the flux/telluric correction algorithm of \citet{2003PASP..115..389V}.

\subsection{High Angular Resolution Imaging}\label{sec:gemini}


We performed high-resolution speckle imaging observations to search for close-in unresolved stellar companions.


We observed TOI-5575 on UT 2023 May 6 with the speckle polarimeter on the 2.5-m telescope at the Caucasian Observatory of Sternberg Astronomical Institute (SAI) of Lomonosov Moscow State University. A low--noise CMOS detector Hamamatsu ORCA--quest \citep{Strakhov2023} was used as a detector. The atmospheric dispersion compensator was active, which allowed using the $I_\mathrm{c}$ band. The respective angular resolution is $0.083^{\prime\prime}$. No companion was detected. The detection limits at distances $0.25$ and $1.0^{\prime\prime}$ from the star are $\Delta I_\mathrm{c}=2.2$ and $4.5$ magnitudes, see Figure~\ref{fig:spp}.


TOI-5575 was also observed on UT 2025 April 14 using the ‘Alopeke speckle instrument on the Gemini North 8-m telescope \citep{Scott2021}. ‘Alopeke provides simultaneous speckle imaging in two filters with output data products including reconstructed images and robust contrast limits on companion detections \citep{Howell2016}. Ten sets of 1000 $\times$ 0.06 sec exposures were collected for TOI-5575 at 562 nm, 832 nm, SDSS $r$, and SDSS $z$, and were reduced following the standard reduction pipeline \citep{Howell2011}. Figure~\ref{fig:spp} shows the final 5$\sigma$ magnitude contrast curves and the 832 nm and SDSS $z$-band reconstructed images. There are hints of a close-in companion in the reconstructed $z$-band image, the signal of which is likely caused by imaging smearing and significant chromatic effects due to the high airmass \citep{Clark2024AJ....167...56C}. Furthermore, TOI-5575 has a RUWE value of 1.2, which is consistent with a single star \citep{Ziegler2020AJ....159...19Z, Gaia2021A&A...649A...1G}. We also find no RV linear trend in our SPIRou data (see Section~\ref{jointfit}). While we cannot rule out the presence of a close-in ($<1''$) companion entirely, we find it unlikely to be real for the reasons above. No other companion brighter than 4-5 magnitudes was detected down to the diffraction limit of the telescope (20 mas) out to 1.2$''$ (1.1 to 68 au).


\section{Analysis}\label{Analysis}

\subsection{Stellar Characterization}

Figure~\ref{fig:sed} displays the reduced NIRES spectrum of TOI-5575 with telluric absorption regions masked, compared to similar data of the best-matching M5 standard Gliese 866AB from \citet{2005ApJ...623.1115C}. We use the relations of \citet{2014AJ....147..160M} to infer [Fe/H] = $-$0.21$\pm$0.07 based on a combination of H$_2$O-K2 color and 2.20~$\mu$m Na I and 2.26~$\mu$m Ca I equivalent widths. We infer equivalent subsolar metallicities based on near-infrared relations defined in \citet{2012ApJ...748...93R,2012ApJ...747L..38T}; and \citet{2014AJ....147...20N}.

We next estimate the stellar properties based on empirical relations. We compute the absolute magnitude $M_{K}=7.69\pm0.02$ of \tar\ based on the $m_{K}$ from 2MASS and parallax from \gaia\ DR3. By utilizing the polynomial relation between $R_{\ast}$ and $M_{K}$ derived in \cite{Mann2015}, we obtain a stellar radius of $\rm 0.24\pm0.02\,R_\odot$. In addition, we find a stellar mass of $\rm 0.21\pm0.02\,M_\odot$ according to the $M_{\ast}$-$M_{K}$ relation \citep{Mann2019}, which agrees well with the result of $\rm 0.21\pm0.02\,M_\odot$ from \cite{Benedict2016}. Finally, the stellar effective temperature $T_{\rm eff}$ is estimated to be $3115\pm100$\,K through a polynomial relation with stellar colors $V-J$ and $J-H$ \citep{Mann2015}, in accord with $3029\pm120$\,K from \citep{Pecaut2013}, which is also based on an empirical equation but between $T_{\rm eff}$ and $V-K$. 

We also independently determine the stellar radius and effective temperature of \tar\ by carrying out a spectral energy distribution (SED) analysis \citep{Stassun:2016,Stassun:2017,Stassun:2018}. We pull the $JHK$ magnitudes from {\it 2MASS} \citep{Cutri2003}, the $W1$, $W2$ and $W3$ magnitudes from {\it WISE} \citep{wright2010} as well as three {\it Gaia} magnitudes $G$, $G_{\rm BP}, G_{\rm RP}$ from {\it Gaia} DR3 \citep{Gaia2023}, all listed in Table~\ref{starparam}. Together, the available photometry spans the wavelength range 0.5--10~$\mu$m (Figure~\ref{fig:sed}). We perform a fit using PHOENIX stellar atmosphere models \citep{Husser2013}, with the free parameters being the effective temperature ($T_{\rm eff}$) and metallicity ([Fe/H]). Given the proximity of the host star, we fix the extinction $A_V$ to 0. The best-fit has a reduced $\chi^2$ of 1.8 with $T_{\rm eff} = 3175 \pm 100$\,K and [Fe/H] = $-0.3 \pm 0.5$~dex, consistent with the result we obtained from the NIRES spectrum. The metallicity is further constrained by the {\it Gaia\/} spectrum (see gray swathe in the inset of Figure~\ref{fig:sed}). Integrating the (unreddened) model SED gives the bolometric flux at Earth, $F_{\rm bol} = 5.32 \pm 0.19 \times 10^{-11}$ erg~s$^{-1}$~cm$^{-2}$. Taking the $F_{\rm bol}$ and $T_{\rm eff}$ together with the {\it Gaia\/} DR3 parallax leads to the stellar radius R$_\ast = 0.24 \pm 0.02$\,R$_\odot$, consistent with the estimation above. The final stellar parameters adopted for further analysis are listed in Table~\ref{starparam}. 




\subsection{Joint Fit of Transit and RV data}\label{jointfit}

We utilize the \code{juliet} \citep{juliet} package to perform a joint fit to the space and ground-based photometry together with the SPIRou radial velocity data in order to determine the physical parameters of the companion. To summarize, \code{juliet} makes use of \code{batman} \citep{Kreidberg2015} to fit the light curves and \code{radvel} \citep{Fulton2018} to build the standard RV model. We apply the dynamic nested sampling method with \code{dynesty} \citep{Speagle2019} to obtain the posteriors of each parameter. 


Since the TESS images taken during the Primary, First Extended and Second Extended Missions have different cadences, we treat the data as if they were from three different instruments in the joint fit. Due to the relatively large pixel scale ($21''\times 21''$) of TESS, we fit dilution factors\footnote{The dilution factor defined in the \code{juliet} package is given by 1/(1+$A_{D}$), where $A_{D}$ is the flux ratio between contaminating sources and the target.} for the TESS photometry to account for the potential light contamination from other stars. We fix it at 1 for the ground-based ASP data because of its small pixel scale that deblends all nearby stars. We adopt the quadratic limb darkening law \cite[$q_{1}$ and $q_{2}$;][]{Kipping2013} for both TESS and ASP data. Moreover, we also model the relative out-of-transit target flux for each photometric dataset and incorporate a jitter term to take additional white noises into account. 

Regarding the RV modeling, the short time span and the limited number of our SPIRou observations make it insufficient to search for Keplerian signals from additional outer planets other than the one from \tar b, hence we focus exclusively on the Keplerian model of a single companion and fix the linear and quadratic RV trends at zero\footnote{An independent fit including a linear RV trend leads to a best-fit of $\dot{\gamma}=1.1\pm4.4$ $\rm m\ s^{-1}\ day^{-1}$, consistent with 0 within $1\sigma$.}. As the companion has a long orbital period and tidal effects are weak (see Section~\ref{discussion_toi5575}), we fit an eccentric orbit, leaving $e\sin \omega$ and $e\cos \omega$ as free parameters. We also model a constant RV offset and consider possible additional systematics by fitting an RV jitter, which is added in quadrature to the original RV uncertainties. We allow all parameters to vary uniformly or loguniformly except for the orbital period and mid-transit time, on which we place narrow Gaussian normal priors, centering at the results from transit search with $1\sigma$ values of 0.1 days. All photometric and RV data along with the best-fit transit and Keplerian models are illustrated in Figure~\ref{tess_joint_fit}. The priors we set and the resulting best-fit key parameters are summarized in Table~\ref{allposteriors}.

We next perform a secondary eclipse analysis by masking out the secondary eclipse data based on the joint fit and then detrending the light curve with GP. We do not find any secondary eclipse signal in the TESS light curve with depth larger than 1.3\% ($3\sigma$ confidence level). Supposing that the companion is a very-low-mass M dwarf with $M_\ast=0.08\ {\rm M_\odot}$, the expected secondary eclipse depth in the TESS spectral bandpass would be about 2.3\%, assuming blackbody spectra without reflected starlight \citep{Charbonneau2005}. Consequently, we conclude that the companion is a massive brown dwarf rather than a star at the bottom of the main sequence. We further carry out a transit timing variation (TTV) analysis by modeling the mid-transit time of individual transits of TESS and ASP data, and fit a linear ephemeris to the transit times using \code{juliet}. The results show that all timing variations are within 4 minutes with no significant TTV signal detected in the dataset. 



\begin{table*}\scriptsize
    {\renewcommand{\arraystretch}{1.05}
    \caption{Priors and the best-fit values along with the 68\% credibility intervals in the final joint fit for \tar. $\mathcal{N}$($\mu\ ,\ \sigma^{2}$) is a normal prior with mean $\mu$ and standard deviation $\sigma$. $\mathcal{U}$(a\ , \ b) represents a uniform prior between $a$ and $b$. $\mathcal{LU}$(a\ , \ b) stands for a log-uniform prior between $a$ and $b$.}
    \begin{tabular}{lccr}
        \hline\hline
        Parameter       &Prior &Value    &Description\\\hline
        \it{Orbit parameters}\\
        $P$ (days)  &$\mathcal{N}$ ($32.1$\ ,\ $0.1$)  &$32.072306\pm0.000007$
        &Orbital period.\\
        $T_{c}$ (BJD-2457000) &$\mathcal{N}$ ($1689.2$\ ,\ $0.1$) &$1689.2666\pm0.0003$ &Mid-transit time.\\
        $r_1$ &$\mathcal{U}$ ($0$\ ,\ $1$) &$0.6391\pm0.0015$ &Parametrization for p and b$^{[1]}$.\\
        $r_2$ &$\mathcal{U}$ ($0$\ ,\ $1$) &$0.3518\pm0.0065$ &Parametrization for p and b.\\
        $e\sin \omega$ &$\mathcal{U}$ ($-1$\ ,\ $1$) &$0.0338\pm0.0030$ &Parametrization for $e$ and $\omega$.\\
        $e\cos \omega$ &$\mathcal{U}$ ($-1$\ ,\ $1$) &$-0.1838\pm0.0020$ &Parametrization for $e$ and $\omega$.\\
        $K$ (m~s$^{-1}$) &$\mathcal{U}$ ($5000$\ ,\ $15000$) &$11041.6\pm23.5$ &RV semi-amplitude.\\
        $\mu_{\rm SPIRou}$ (m~s$^{-1}$) &$\mathcal{U}$ ($-50000$\ ,\ $-30000$) &$-45931.7\pm 32..2$ &RV offset.\\
        $\sigma_{\rm SPIRou}$ (m~s$^{-1}$) &$\mathcal{U}$ ($0$\ ,\ $100$) &$49.8\pm 9.2$ &RV jitter.\\\hline
        \it{Stellar parameter}\\
        $\rho_{\ast}$ (kg~m$^{-3}$) &$\mathcal{LU}$ ($10^{2}$\ ,\ $10^{5}$) &$21500\pm121$ &Stellar density of the host star.\\\hline
        \it{Dilution factors}\\
        $D_{\rm TESS, PM}^{[2]}$ &$\mathcal{U}$ ($0$\ ,\ $1$) &$ 0.997\pm0.002$ & \\
        $D_{\rm TESS, EM1}$ &$\mathcal{U}$ ($0$\ ,\ $1$) &$0.838\pm0.014$ & \\
        $D_{\rm TESS, EM2}$ &$\mathcal{U}$ ($0$\ ,\ $1$) &$0.905\pm0.011$ & \\
        $D_{\rm ASP}$ &1 (Fixed) &$\cdots$ & \\\hline
        \it{Derived parameters}\\
        $R_{c}/R_{\ast}$ &$\cdots$ &$0.3518\pm0.0065$ &Scaled companion radius.\\
        $a/R_{\ast}$ &$\cdots$ &$105.35\pm0.21$ &Scaled semi-major axis.\\
        $b$ &$\cdots$ &$0.459\pm0.022$ &Impact parameter.\\
        $i$ (degrees) &$\cdots$ &$89.73\pm0.02$ &Orbital inclination.\\
        $e$ &$\cdots$ &$0.187\pm0.002$ &Orbital eccentricity.\\
        $\omega$ (degrees) &$\cdots$ &$169.6\pm1.1$ &Argument of periapsis.\\\hline
        \it{Physical parameters of the BD}\\
        $R_{c}$ ($\rm R_{J}$) &$\cdots$ &$0.84\pm 0.07$ &Companion radius.\\
        $M_{c}$ ($\rm M_{J}$) &$\cdots$ &$72.4\pm 4.1$ &Companion mass.\\
        $a$ (AU) &$\cdots$ &$0.118\pm 0.010$ &Semi-major axis.\\
        $q$ &$\cdots$ &$0.329\pm 0.037$ &Mass ratio between the BD and the host star.\\
        \hline
    \label{allposteriors}    
    \end{tabular}}
    \begin{tablenotes} 
       \item[1]  [1] The parameters of p and b are the companion-to-star radius ratio and the impact parameter \citep{Espinoza2018}. [2] The subscripts PM, EM1 and EM2 represent Primary Mission, First Extended Mission and Second Extended Mission. The light dilution factors are different because of the different sources of TESS photometry based on different apertures.
    \end{tablenotes}
\end{table*}

\subsection{TOI-5575b: A Massive Eccentric Brown Dwarf Around a Low-Mass M Star}\label{discussion_toi5575}

\begin{figure*}
\centering
\includegraphics[width=0.49\textwidth]{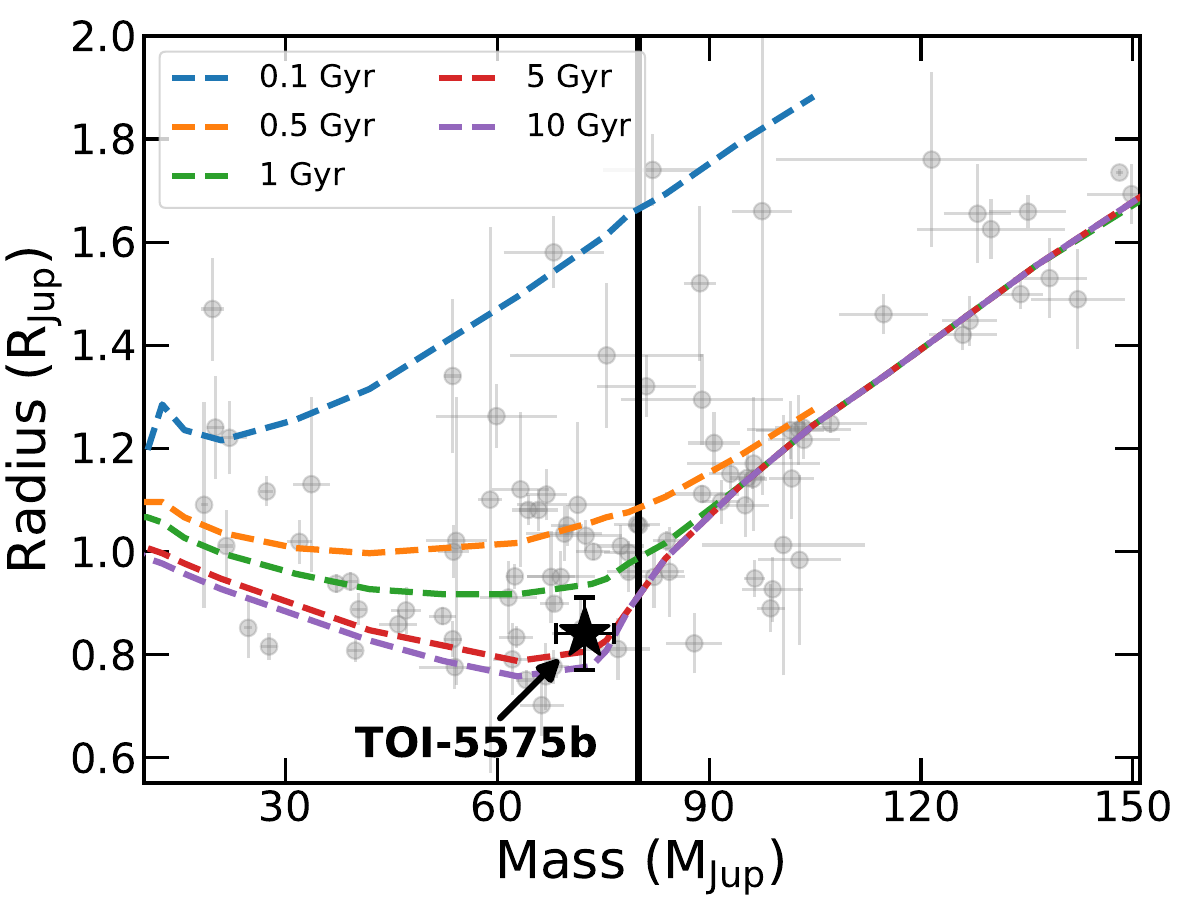}
\includegraphics[width=0.49\textwidth]{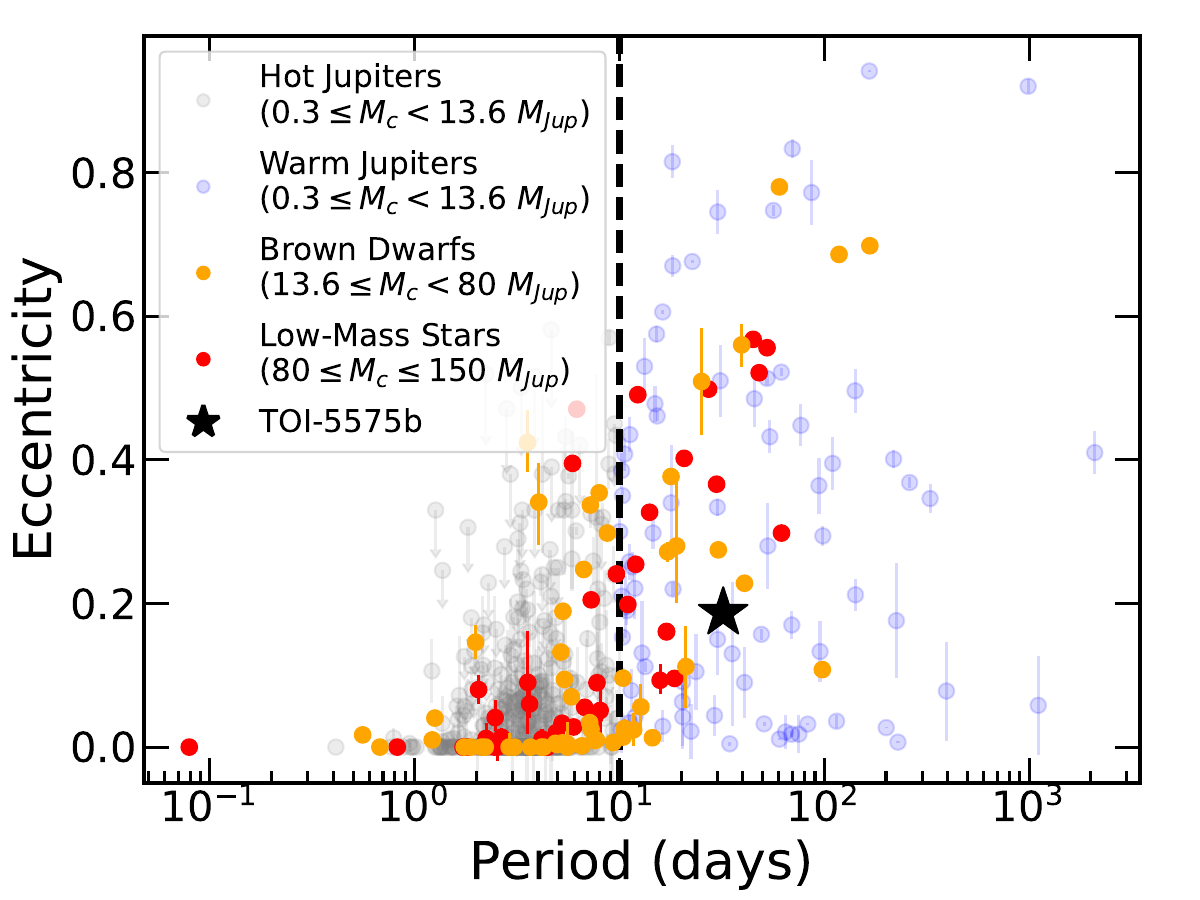}
\caption{{\it Left panel:} The mass and radius diagram of all confirmed transiting brown dwarfs and low-mass stars. The colored dashed lines are the theoretical isochrones of cool brown dwarfs with different ages at solar metallicity from \cite{Baraffe2003}. The vertical solid line marks the mass of $\rm 80\ M_{J}$. {\it Right panel:} The eccentricity vs. orbital periods of known transiting brown dwarfs (orange) and low-mass stars (red). The vertical dashed line marks the 10-day period boundary that we used to select long-period objects for statistics. The blue dots are the selected transiting warm Jupiters while the transiting hot Jupiters are shown as gray dots for reference. In both panels, \tar b is shown as a black star.}
\label{evolution_track}
\end{figure*}

The joint photometric and RV analysis reveals that the companion is a brown dwarf with a radius of $0.84\pm0.07$\, R$_{\rm J}$ and a mass of $72.4\pm4.1$\,M$_{\rm J}$ on a 32-day eccentric ($e=0.187\pm0.002$) orbit. The location of \tar b in the Mass-Radius diagram (Figure~\ref{evolution_track}) aligns with the evolution track of a brown dwarf with an age of about 5-10 Gyr \citep{Baraffe2003}, suggesting that the host star is not young, which agrees with the non-detection of significant rotation signal in the light curve from Zwicky Transient Facility \citep[ZTF;][]{Bellm2019,Masci2019}\footnote{A total of two ZTF light curves in $g$ and $r$ bands are available and each has 475 data points, spanning about 5.5 years.}. Following Equation 1 in \cite{Jackson2008}, we estimate the tidal circularization timescale $\tau_{circ}$ of \tar b. By adopting the typical modified quality factor of the brown dwarf ($Q_{BD}$) and the host star ($Q_\ast$) both as $10^{6}$, we obtain a $\tau_{circ}$ of about $5\times10^{13}$~years, much longer than any astrophysical timescale. Therefore, we conclude that \tar b holds the primordial configuration memory of the eccentricity without being erased by tidal effects. With a period of about 32 days, \tar b joins the small group of long-period sub-stellar objects with $P>10$~days (see the right panel of Figure~\ref{evolution_track}).

\begin{figure*}
\centering
\includegraphics[width=0.99\textwidth]{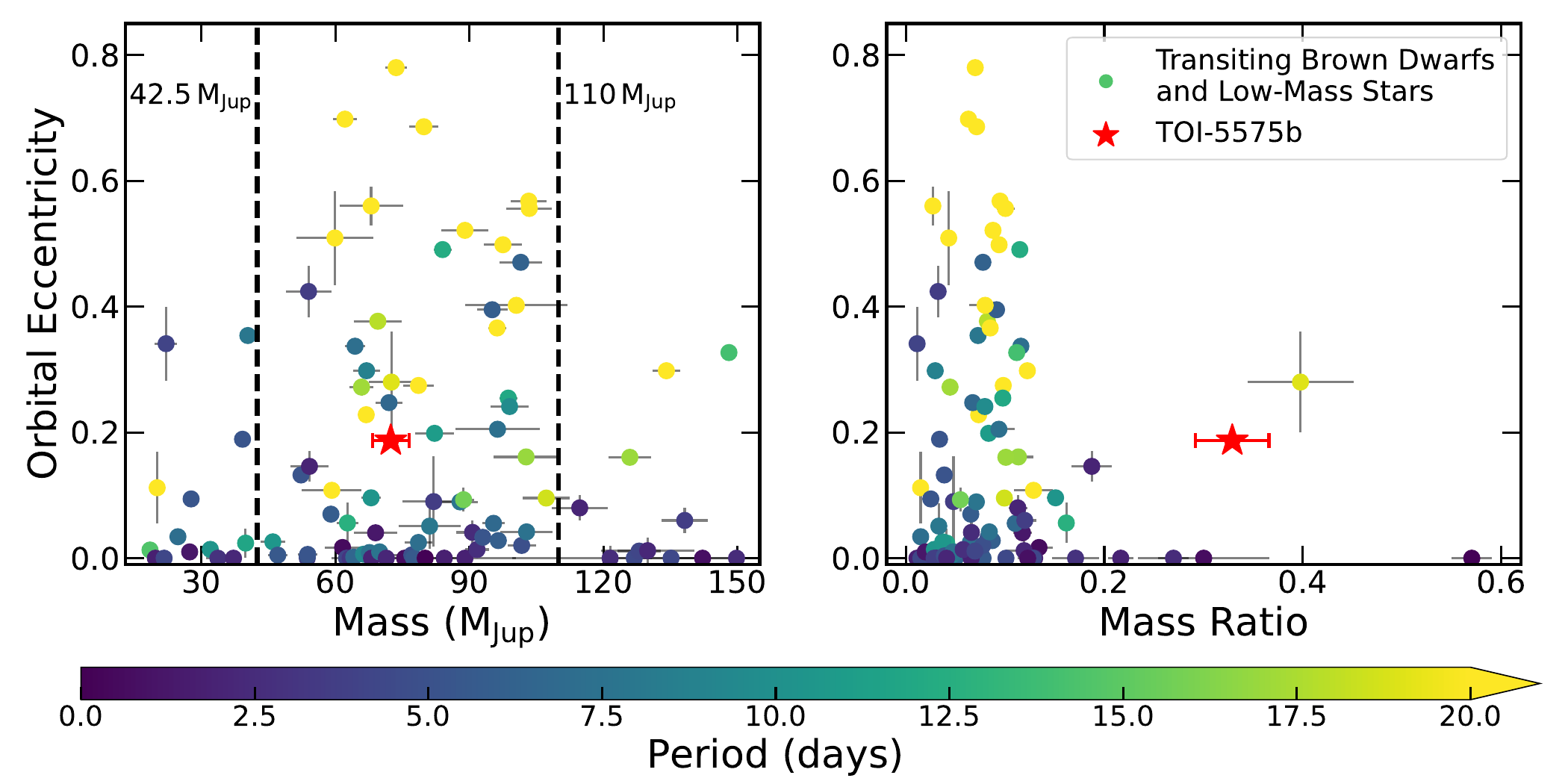}
\caption{{\it Left panel:} The eccentricity and mass diagram of all confirmed transiting brown dwarfs and low-mass stars colored by the orbital period. Two vertical black dashed lines mark the mass range ($42.5\leq M_c\leq 110\,{\rm M_J}$) where the companions are more likely to have large eccentricities (see Section~\ref{discussion_toi5575} for details). {\it Right panel:} Same to the left but for the companion-to-host mass ratio. In both panels, \tar b is shown as a red star.} 
\label{ecc_mass_massratio}
\end{figure*}

\tar b is so far the third highest mass ratio transiting brown dwarf system ($q=0.329\pm0.037$) following ZTF~J2020+5033b \citep{ElBadry2023} and TOI-6508b \citep{Barkaoui2025}. Its moderate eccentricity places it among the rare sample of eccentric-orbit brown dwarfs. Figure~\ref{ecc_mass_massratio} presents the eccentricity-mass diagram of known transiting brown dwarfs and low-mass stars with companion masses $M_c\leq 150\,{\rm M_J}$. Two key features can be seen: 1) the majority of eccentric ($e>0.3$) companions occupy the mass range between $\rm 42.5\,M_J$ and $\rm 110\,M_J$; 2) most of them have orbital periods longer than 10 days. The outer mass threshold of $110\,{\rm M_J}$ is simply from visual inspection while the inner mass limit of $\rm 42.5\,M_J$ was initially reported by \cite{Ma2014}, who found that brown dwarfs with masses above and below $\rm 42.5\,M_J$ have different eccentricity distributions, indicating two different formation mechanisms. We compute the number ratios of high-eccentricity objects with $e>0.3$ within three mass ranges: $M_c<42.5\ {\rm M_J}$, $42.5 \leq M_c<110\ {\rm M_J}$ and $M_c\geq 110\ {\rm M_J}$. We find that the fractions are 14.3\%, 24.6\% and 8.3\%, showing a tentative increase of high-eccentricity objects with $42.5 \leq M_c<110\ {\rm M_J}$. However, \cite{Vowell2025} argued against a transition point in the eccentricity distribution when considering the systems' mass ratio (see the right panel of Figure~\ref{ecc_mass_massratio}), a more fundamental parameter of the formation process, where the eccentricity dichotomy seems to disappear. 


\section{Eccentricity Distribution}\label{Eccentricity_Distribution}

In this section, we compare the eccentricity distribution of transiting long-period giant planets, brown dwarfs as well as low-mass stars. We first outline how we build the samples in Section~\ref{sample_selection} and then describe the way we model the eccentricity distributions before summarizing the results in Section~\ref{model_ecc_dis}.

\subsection{Sample Selection}\label{sample_selection}

We construct our brown dwarf and low-mass star samples based on a combination of two catalogs published in \cite{Vowell2025} and \cite{Barkaoui2025}, supplemented with several recent detections \citep{Larsen2025,Zhang2025} as well as the newly discovered \tar b system reported in this work. We emphasize that all companions in the list are transiting systems, and therefore have true mass measurements. We limit our brown dwarf population to companions with masses between 13.6 and 80\,$\rm M_{J}$, and filter out low-mass star systems that have companion masses $80\leq M_{c}\leq 150\,{\rm M_J}$. We exclude short-period systems and only accept objects with $10\leq P\lesssim 1000$~days (corresponding to a semi-major axis range about 0.1-1.5 AU) to minimize the impact of stellar tidal effect that could erase the orbital eccentricity (see the eccentricity and period diagram in Figure~\ref{evolution_track}). The inner period boundary is selected based on the tidal circularization timescale $\tau_{circ}$. Assuming that the modified quality factors of a brown dwarf ($Q_{BD}$) and its host star ($Q_\ast$) both as typical values $10^{6}$ \citep{Goldreich1966,Jackson2008}, a brown dwarf with mass between 13-80\,$\rm M_J$ and a period of 10 days orbiting a Sun-like star has a tidal circularization timescale $\tau_{circ}$ between 18 and 84 Gyrs using Equation 1 in \cite{Jackson2008}, which we treat as unaffected by tidal effects. Eventually, a total of 19 brown dwarfs and 15 low-mass stars are left in our sample. We list all their properties in Table~\ref{catalog}.

\begin{table*}\scriptsize
    \centering
    {\renewcommand{\arraystretch}{1.0}
    \caption{Properties of the transiting long-period brown dwarf and low-mass star sample$^{[1]}$}
    \begin{tabular}{lcccccccc}
        \hline\hline
        Object &Period [d] &$M_{2}$ [$\rm M_J$] &$R_{2}$ [$\rm R_J$] &Eccentricity &$M_{1}$ [$\rm M_\odot$] &$R_{1}$ [$\rm R_\odot$] &$T_{\rm eff}$ [K] &[Fe/H] \\\hline
        \it{Brown Dwarfs {\rm (19 in total)}: }\\
        TOI-1278b &14.48 &$18.50_{-0.50}^{+0.50}$ &$1.09_{-0.20}^{+0.24}$ &$0.013_{-0.004}^{+0.004}$ &$0.54_{-0.02}^{+0.02}$ &$0.57_{-0.01}^{+0.01}$ &$3799_{-42}^{+42}$ &$-0.01_{-0.28}^{+0.28}$ \\
Kepler-39b &21.09 &$20.10_{-1.20}^{+1.30}$ &$1.24_{-0.10}^{+0.09}$ &$0.112_{-0.057}^{+0.057}$ &$1.29_{-0.07}^{+0.06}$ &$1.40_{-0.10}^{+0.10}$ &$6350_{-100}^{+100}$ &$0.10_{-0.14}^{+0.14}$ \\
TOI-4776b &10.41 &$32.00_{-1.80}^{+1.90}$ &$1.02_{-0.04}^{+0.05}$ &$0.014_{-0.010}^{+0.023}$ &$1.06_{-0.07}^{+0.07}$ &$1.22_{-0.40}^{+0.46}$ &$6011_{-33}^{+33}$ &$-0.05_{-0.05}^{+0.05}$ \\
KOI-205b$^{[2]}$ &11.72 &$39.90_{-1.00}^{+1.00}$ &$0.81_{-0.02}^{+0.02}$ &$0.024_{-0.023}^{+0.023}$ &$0.93_{-0.03}^{+0.03}$ &$0.84_{-0.02}^{+0.02}$ &$5237_{-60}^{+60}$ &$0.14_{-0.12}^{+0.12}$ \\
TOI-1406b &10.57 &$46.00_{-2.70}^{+2.60}$ &$0.86_{-0.03}^{+0.03}$ &$0.026_{-0.010}^{+0.013}$ &$1.18_{-0.09}^{+0.08}$ &$1.35_{-0.03}^{+0.03}$ &$6290_{-100}^{+100}$ &$-0.08_{-0.09}^{+0.09}$ \\
RIK 72b &97.76 &$59.20_{-6.70}^{+6.80}$ &$3.10_{-0.31}^{+0.31}$ &$0.108_{-0.006}^{+0.012}$ &$0.44_{-0.04}^{+0.04}$ &$0.96_{-0.10}^{+0.10}$ &$3349_{-142}^{+142}$ &$0.00_{-0.10}^{+0.10}$ \\
TOI-811b &25.17 &$59.90_{-8.60}^{+13.00}$ &$1.26_{-0.06}^{+0.06}$ &$0.509_{-0.075}^{+0.075}$ &$1.32_{-0.07}^{+0.05}$ &$1.27_{-0.09}^{+0.06}$ &$6107_{-77}^{+77}$ &$0.40_{-0.09}^{+0.07}$ \\
KOI-415b &166.79 &$62.14_{-2.69}^{+2.69}$ &$0.79_{-0.07}^{+0.12}$ &$0.698_{-0.002}^{+0.002}$ &$0.94_{-0.06}^{+0.06}$ &$1.25_{-0.10}^{+0.15}$ &$5810_{-80}^{+80}$ &$-0.24_{-0.11}^{+0.11}$ \\
LHS 6343b &12.71 &$62.70_{-2.40}^{+2.40}$ &$0.83_{-0.02}^{+0.02}$ &$0.056_{-0.032}^{+0.032}$ &$0.37_{-0.01}^{+0.01}$ &$0.38_{-0.01}^{+0.01}$ &$3130_{-20}^{+20}$ &$0.04_{-0.08}^{+0.08}$ \\
TOI-1982b &17.17 &$65.85_{-2.72}^{+2.75}$ &$1.08_{-0.04}^{+0.04}$ &$0.272_{-0.014}^{+0.014}$ &$1.41_{-0.08}^{+0.08}$ &$1.51_{-0.05}^{+0.05}$ &$6325_{-110}^{+110}$ &$-0.10_{-0.09}^{+0.09}$ \\
EPIC 201702477b &40.74 &$66.90_{-1.70}^{+1.70}$ &$0.76_{-0.07}^{+0.07}$ &$0.228_{-0.003}^{+0.003}$ &$0.87_{-0.03}^{+0.03}$ &$0.90_{-0.06}^{+0.06}$ &$5517_{-70}^{+70}$ &$-0.16_{-0.05}^{+0.05}$ \\
HIP 33609b &39.47 &$68.00_{-7.10}^{+7.40}$ &$1.58_{-0.07}^{+0.07}$ &$0.560_{-0.031}^{+0.029}$ &$2.38_{-0.10}^{+0.10}$ &$1.86_{-0.08}^{+0.09}$ &$10400_{-660}^{+800}$ &$-0.01_{-0.20}^{+0.19}$ \\
TOI-5389Ab &10.40 &$68.00_{-2.20}^{+2.20}$ &$0.78_{-0.03}^{+0.04}$ &$0.096_{-0.005}^{+0.003}$ &$0.43_{-0.02}^{+0.02}$ &$0.42_{-0.02}^{+0.02}$ &$ 3569_{-59}^{+59}$ &$-0.15_{-0.16}^{+0.16}$\\
NGTS-19b &17.84 &$69.50_{-5.40}^{+5.70}$ &$1.03_{-0.05}^{+0.06}$ &$0.377_{-0.006}^{+0.006}$ &$0.81_{-0.04}^{+0.04}$ &$0.90_{-0.04}^{+0.04}$ &$4716_{-28}^{+39}$ &$0.11_{-0.07}^{+0.07}$ \\
\textbf{TOI-5575b (This Work)} &32.07 &$72.40_{-4.10}^{+4.10}$ &$0.84_{-0.07}^{+0.07}$ &$0.187_{-0.002}^{+0.002}$ &$0.21_{-0.02}^{+0.02}$ &$0.24_{-0.02}^{+0.02}$ &$3115_{-100}^{+100}$ &$-0.21_{-0.07}^{+0.07}$ \\
TOI-6508b &18.99 &$72.53_{-5.09}^{+7.61}$ &$1.03_{-0.03}^{+0.03}$ &$0.280_{-0.080}^{+0.090}$ &$0.17_{-0.02}^{+0.02}$ &$0.20_{-0.01}^{+0.01}$ &$3003_{-69}^{+71}$ &$-0.22_{-0.08}^{+0.08}$ \\
TOI-2490b &60.33 &$73.60_{-2.40}^{+2.40}$ &$1.00_{-0.02}^{+0.02}$ &$0.7799_{-0.0005}^{+0.0005}$ &$1.00_{-0.02}^{+0.03}$ &$1.10_{-0.01}^{+0.01}$ &$5558_{-80}^{+80}$ &$0.32_{-0.05}^{+0.05}$ \\
KOI-189b &30.36 &$78.60_{-3.50}^{+3.50}$ &$1.00_{-0.02}^{+0.02}$ &$0.275_{-0.004}^{+0.004}$ &$0.76_{-0.05}^{+0.05}$ &$0.73_{-0.02}^{+0.02}$ &$4952_{-40}^{+40}$ &$-0.12_{-0.10}^{+0.10}$ \\
Kepler-807b &117.93 &$79.80_{-3.30}^{+3.40}$ &$1.05_{-0.49}^{+0.51}$ &$0.686_{-0.002}^{+0.002}$ &$1.07_{-0.07}^{+0.07}$ &$1.20_{-0.04}^{+0.05}$ &$5994_{-78}^{+78}$ &$0.04_{-0.05}^{+0.06}$ \\     \hline 
       \it{Low-Mass Stars {\rm (15 in total)}: }\\
        TOI-746B &10.98 &$82.20_{-4.40}^{+4.90}$ &$0.95_{-0.06}^{+0.09}$ &$0.199_{-0.003}^{+0.003}$ &$0.94_{-0.08}^{+0.09}$ &$0.97_{-0.03}^{+0.04}$ &$5690_{-140}^{+140}$ &$-0.02_{-0.23}^{+0.23}$ \\
TOI-4635B &12.28 &$84.00_{-2.00}^{+2.10}$ &$1.02_{-0.02}^{+0.02}$ &$0.491_{-0.002}^{+0.002}$ &$0.70_{-0.03}^{+0.03}$ &$0.68_{-0.01}^{+0.01}$ &$4555_{-74}^{+67}$ &$-0.09_{-0.03}^{+0.04}$ \\
TOI-681B &15.78 &$88.70_{-2.30}^{+2.50}$ &$1.52_{-0.15}^{+0.25}$ &$0.093_{-0.019}^{+0.022}$ &$1.54_{-0.05}^{+0.06}$ &$1.47_{-0.04}^{+0.04}$ &$7440_{-140}^{+150}$ &$-0.08_{-0.05}^{+0.05}$ \\
TOI-694B &48.05 &$89.00_{-5.30}^{+5.30}$ &$1.11_{-0.02}^{+0.02}$ &$0.521_{-0.002}^{+0.002}$ &$0.97_{-0.04}^{+0.05}$ &$1.00_{-0.01}^{+0.01}$ &$5496_{-81}^{+87}$ &$0.21_{-0.08}^{+0.08}$ \\
TIC 320687387B &29.77 &$96.20_{-2.00}^{+1.90}$ &$1.14_{-0.02}^{+0.02}$ &$0.366_{-0.003}^{+0.003}$ &$1.08_{-0.03}^{+0.03}$ &$1.16_{-0.02}^{+0.02}$ &$5780_{-80}^{+80}$ &$0.30_{-0.08}^{+0.08}$ \\
TOI-1213B &27.22 &$97.50_{-4.20}^{+4.40}$ &$1.66_{-0.55}^{+0.78}$ &$0.498_{-0.002}^{+0.003}$ &$0.99_{-0.06}^{+0.07}$ &$0.99_{-0.04}^{+0.04}$ &$5590_{-150}^{+150}$ &$0.25_{-0.14}^{+0.13}$ \\
K2-76B &11.99 &$98.70_{-1.99}^{+1.99}$ &$0.89_{-0.05}^{+0.03}$ &$0.255_{-0.006}^{+0.007}$ &$0.96_{-0.03}^{+0.03}$ &$1.17_{-0.06}^{+0.03}$ &$5747_{-70}^{+64}$ &$0.01_{-0.04}^{+0.04}$ \\
CoRoT 101186644B &20.68 &$100.50_{-11.50}^{+11.50}$ &$1.01_{-0.25}^{+0.06}$ &$0.402_{-0.006}^{+0.006}$ &$1.20_{-0.20}^{+0.20}$ &$1.07_{-0.07}^{+0.07}$ &$6090_{-200}^{+200}$ &$0.20_{-0.20}^{+0.20}$ \\
BD+29 4980B &16.95 &$102.70_{-7.33}^{+7.33}$ &$1.24_{-0.07}^{+0.07}$ &$0.161_{-0.003}^{+0.002}$ &$0.86_{-0.10}^{+0.10}$ &$0.85_{-0.06}^{+0.05}$ &$5150_{-60}^{+90}$ &$0.10_{-0.14}^{+0.14}$ \\
EBLM J2114-39B &44.92 &$103.29_{-3.98}^{+3.98}$ &$1.24_{-0.02}^{+0.02}$ &$0.5677_{-0.0004}^{+0.0005}$ &$1.04_{-0.06}^{+0.06}$ &$1.28_{-0.02}^{+0.02}$ &$5763_{-216}^{+216}$ &$-0.03_{-0.14}^{+0.14}$ \\
KOI-686B &52.51 &$103.40_{-5.10}^{+5.10}$ &$1.22_{-0.04}^{+0.04}$ &$0.556_{-0.004}^{+0.004}$ &$0.98_{-0.07}^{+0.07}$ &$1.04_{-0.03}^{+0.03}$ &$5834_{-100}^{+100}$ &$-0.06_{-0.13}^{+0.13}$ \\
TIC 220568520B &18.56 &$107.20_{-5.20}^{+5.20}$ &$1.25_{-0.02}^{+0.02}$ &$0.096_{-0.003}^{+0.003}$ &$1.03_{-0.04}^{+0.04}$ &$1.01_{-0.01}^{+0.01}$ &$5589_{-81}^{+81}$ &$0.26_{-0.07}^{+0.07}$ \\
EBLM J2343+29B &16.95 &$125.92_{-4.82}^{+4.82}$ &$1.42_{-0.03}^{+0.03}$ &$0.1604_{-0.0003}^{+0.0003}$ &$1.19_{-0.07}^{+0.07}$ &$0.91_{-0.02}^{+0.02}$ &$4984_{-87}^{+87}$ &$0.11_{-0.05}^{+0.05}$ \\
TIC 231005575B &61.78 &$134.10_{-3.14}^{+3.14}$ &$1.50_{-0.03}^{+0.03}$ &$0.298_{-0.001}^{+0.004}$ &$1.04_{-0.04}^{+0.04}$ &$0.99_{-0.05}^{+0.05}$ &$5500_{-85}^{+85}$ &$-0.44_{-0.06}^{+0.06}$ \\
KIC 1571511B &14.02 &$148.10_{-0.53}^{+0.53}$ &$1.74_{-0.01}^{+0.00}$ &$0.327_{-0.003}^{+0.003}$ &$1.26_{-0.03}^{+0.04}$ &$1.34_{-0.01}^{+0.01}$ &$6195_{-50}^{+50}$ &$0.37_{-0.08}^{+0.08}$ \\

        \hline
    \label{catalog}    
    \end{tabular}}
    \begin{tablenotes}
       \item[1] [1] This catalog is built by combining the lists published in \cite{Vowell2025} and \cite{Barkaoui2025} as well as a few recent discoveries \citep{Larsen2025,Zhang2025}. [2]\ The orbital eccentricity of KOI-205b reported in the reference \cite{Diaz2013} is an upper limit ($<0.031$). Here, the result $e=0.024\pm0.023$ is derived based on our independent fit using the same RV dataset. 
    \end{tablenotes}
\end{table*}

Our warm Jupiter sample is retrieved from the NASA Exoplanet Archive\footnote{\url{https://exoplanetarchive.ipac.caltech.edu}} \citep{Akeson2013}. We preselect planets that have radius (i.e., transiting) and eccentricity measurements with mass over $\rm 0.3\,M_J$ but smaller than $\rm 13.6\,M_J$ and the same orbital period boundary as above, leading to 94 systems in total. We threw out 11 warm Jupiters that only have mass upper limits, most of which are mainly due to small RV amplitudes but large RV scatters. We also restrict the sample to systems with determined eccentricities instead of upper limits, and we further exclude 5 systems. In the end, our warm Jupiter sample consists of 78 planets. While the exclusion of objects with eccentricity upper limits could potentially bias toward higher eccentricities, we note that the number of excluded systems is small compared with the total sample size, and that they will not impact our conclusion significantly. 

\subsection{Modeling the Eccentricity Distribution}\label{model_ecc_dis}

We opt to characterize the eccentricity distribution of the transiting long-period giant planet, brown dwarf and low-mass star categories using the Beta distribution, which has been widely used in many previous works \citep[e.g.,][]{Hogg2010,Kipping2013,VanEylen2019,Stevenson2025}. The choice of Beta distribution is motivated by the following two reasons: it is intrinsically defined over the range 0 and 1, and it can produce probability density functions with various shapes using only two parameters. The standard Beta distribution is formulated as:

\begin{equation}
    f(e;\alpha, \beta)= \frac{\Gamma(\alpha+\beta)e^{\alpha-1}(1-e)^{\beta-1}}{\Gamma(\alpha)\Gamma(\beta)},\label{beta_euation}
\end{equation}
where $e$ is the orbital eccentricity, $\Gamma$ represents the gamma function, $\alpha$ and $\beta$ are two parameters that determine the shape of the Beta distribution \citep{Kipping2013}, so-called the shape parameters.

To properly account for measurement uncertainties, we implement a Monte Carlo resampling approach. We resample the eccentricity of each system by randomly drawing samples, assuming truncated Gaussian distributions $\mathcal{TN}(e,\sigma_{e}^{2},0,1)$. Thanks to the large RV amplitudes, the eccentricities of most objects in our sample are precisely determined and the majority of eccentricity uncertainties are at the level of $\lesssim 0.1$, with a median error of 0.023, 0.006 and 0.003 for giant planets, brown dwarfs and low-mass stars, respectively. For those measurements that have different upper and lower errors, we adopt the larger one as $\sigma_{e}$ in order to be conservative. We loop the procedure for 2,000 times for every system, hence we have 2,000 synthetic datasets. We utilize the Beta distribution function embedded in the \code{Scipy} package \citep{Virtanen2020} to perform the fit through the least $\chi^{2}$ method, and record all best-fit shape parameter pairs. We randomly select 30 synthetic datasets and the corresponding best Beta distribution fits, and we validate that these results match with observations.

\begin{figure*}
\centering
\includegraphics[width=0.99\textwidth]{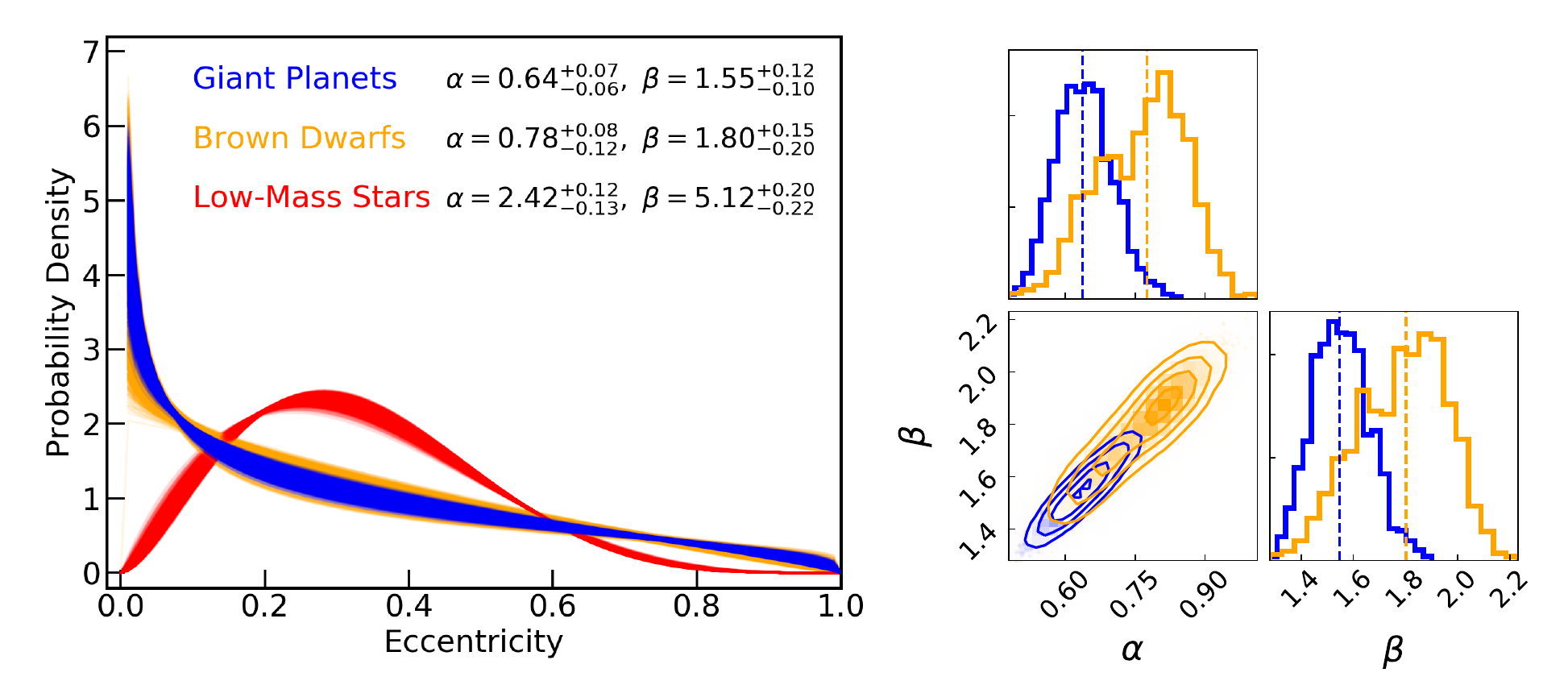}
\caption{{\it Left panel:} The best-fit eccentricity distributions of 2,000 randomly generated eccentricity data sets of transiting long-period ($10 \leq P\lesssim 1000$~days) giant planets (blue, $M_c<13.6\,{\rm M_J}$), brown dwarfs (orange, $13.6\leq M_c<80\,{\rm M_J}$) and low-mass stars (red, $80\leq M_c\leq 150\,{\rm M_J}$), assuming truncated Gaussian distributions (see Section~\ref{model_ecc_dis} for details). The best-fit shape parameters of the Beta distribution (Equation~\ref{beta_euation}) are shown on the top right. {\it Right panel:} The distribution of best-fit shape parameter pairs of 2,000 simulated data sets of giant planets (blue) and brown dwarfs (orange), consistent with each other within about $1\sigma$. The vertical dashed lines show the best-fit parameters. The eccentricity distributions of transiting long-period brown dwarfs are similar to giant planets but different from low-mass stars.} 
\label{betafit}
\end{figure*}

Figure~\ref{betafit} shows the best-fit probability density functions of three companion groups. Giant planets and brown dwarfs with periods longer than 10 days have almost identical eccentricity distributions, preferring circular orbits with an extended tail toward high eccentricities. In the right panel of Figure~\ref{betafit}, we present the resulting shape parameter distributions of all 2,000 simulated datasets of giant planets and brown dwarfs, which overlap with each other within about $1\sigma$. The best-fitting shape parameters of the giant planet sample are $\alpha=0.64_{-0.06}^{+0.07}$ and $\beta=1.55_{-0.10}^{+0.12}$ while the results of brown dwarfs are $\alpha=0.78_{-0.12}^{+0.08}$ and $\beta=1.80_{-0.20}^{+0.15}$. In contrast, low-mass stars within the same period range have a remarkably different behavior ($\alpha=2.42^{+0.12}_{-0.13}$ and $\beta=5.12^{+0.20}_{-0.22}$), which tend to have moderate eccentricities peaking at about 0.3. Finally, by visual inspection, we note that long-period giant planets and brown dwarfs are more likely to have highly eccentric orbits with $e>0.6$ compared with low-mass stars. In fact, all 15 companions in our low-mass star sample have eccentricities below 0.6 while there are 3 of 19 ($15.8\%$) brown dwarfs and 10 of 78 ($12.8\%$) warm Jupiters above this limit. Meanwhile, low-mass stars with circular orbits seem to be rare. The fractions of objects in our sample with $e\leq 0.1$ are 29.5\%, 31.6\% and 13.3\% for giant planets, brown dwarfs and low-mass stars, respectively. 

\begin{figure}
\centering
\subfigure[Samples selected with $a/R_\ast\geq10$]{\includegraphics[width=0.49\textwidth]{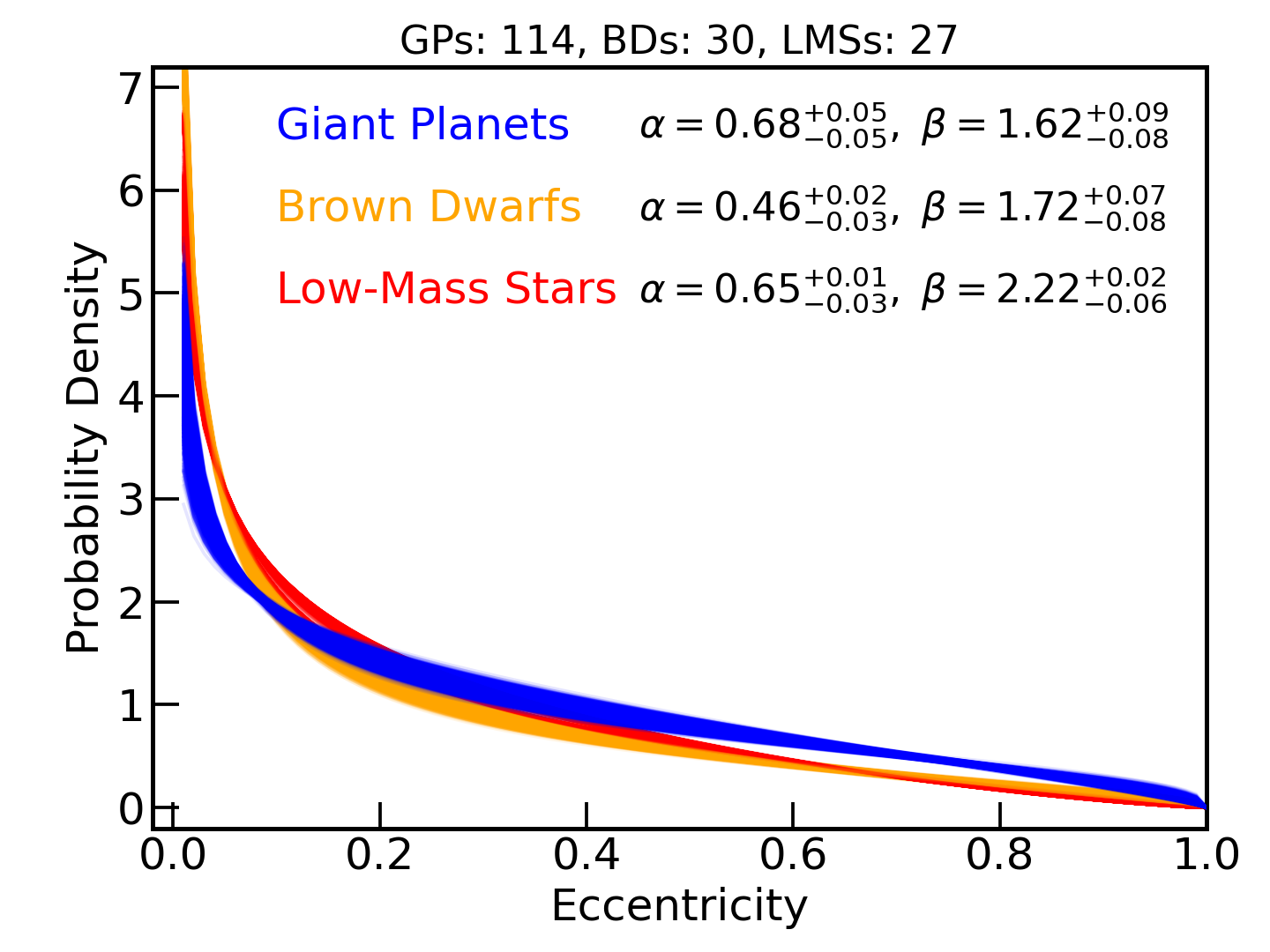}}
\subfigure[Samples selected with $a/R_\ast\geq15$]{\includegraphics[width=0.49\textwidth]{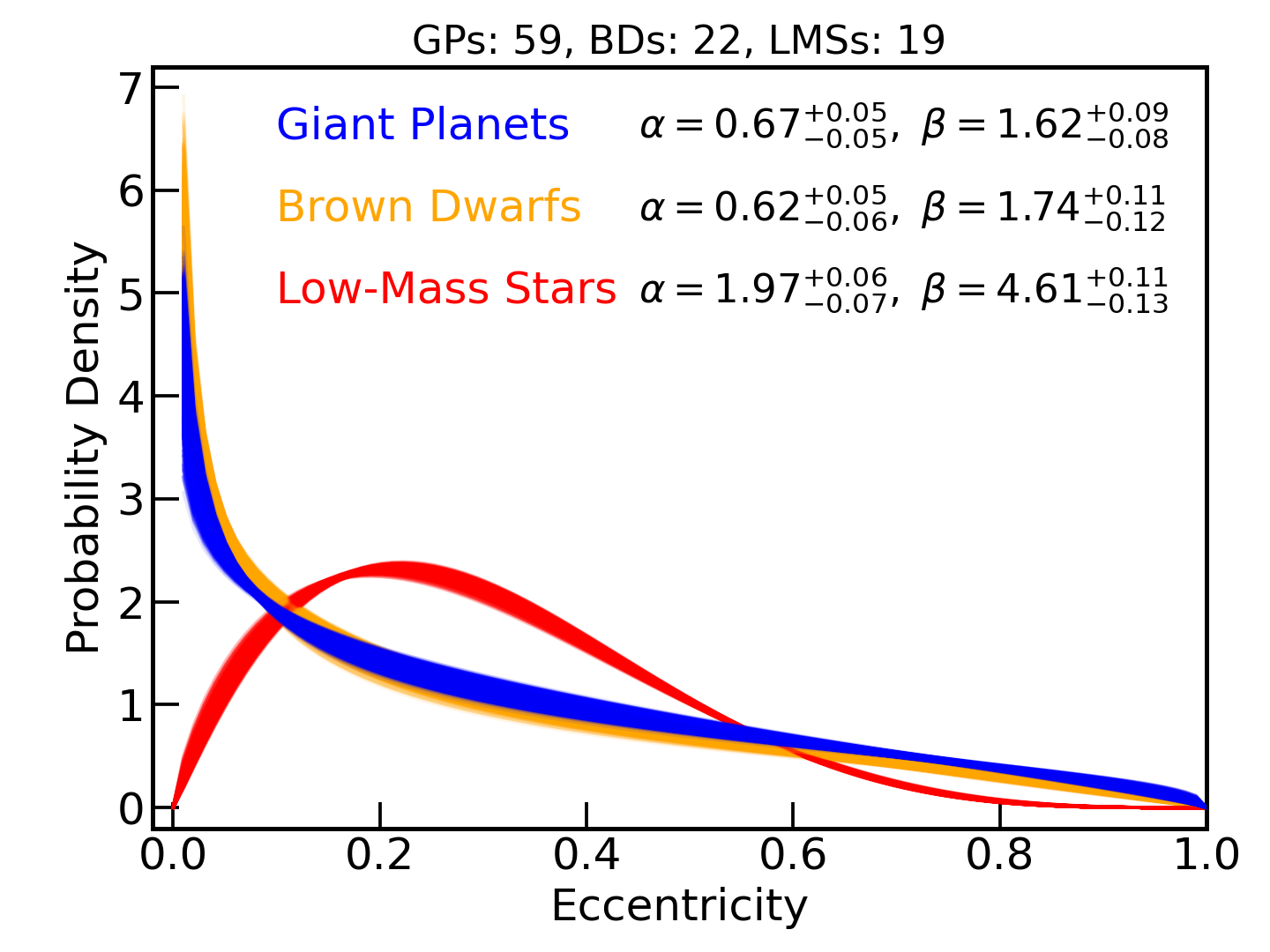}}
\subfigure[Samples selected with $a/R_\ast\geq20$]{\includegraphics[width=0.49\textwidth]{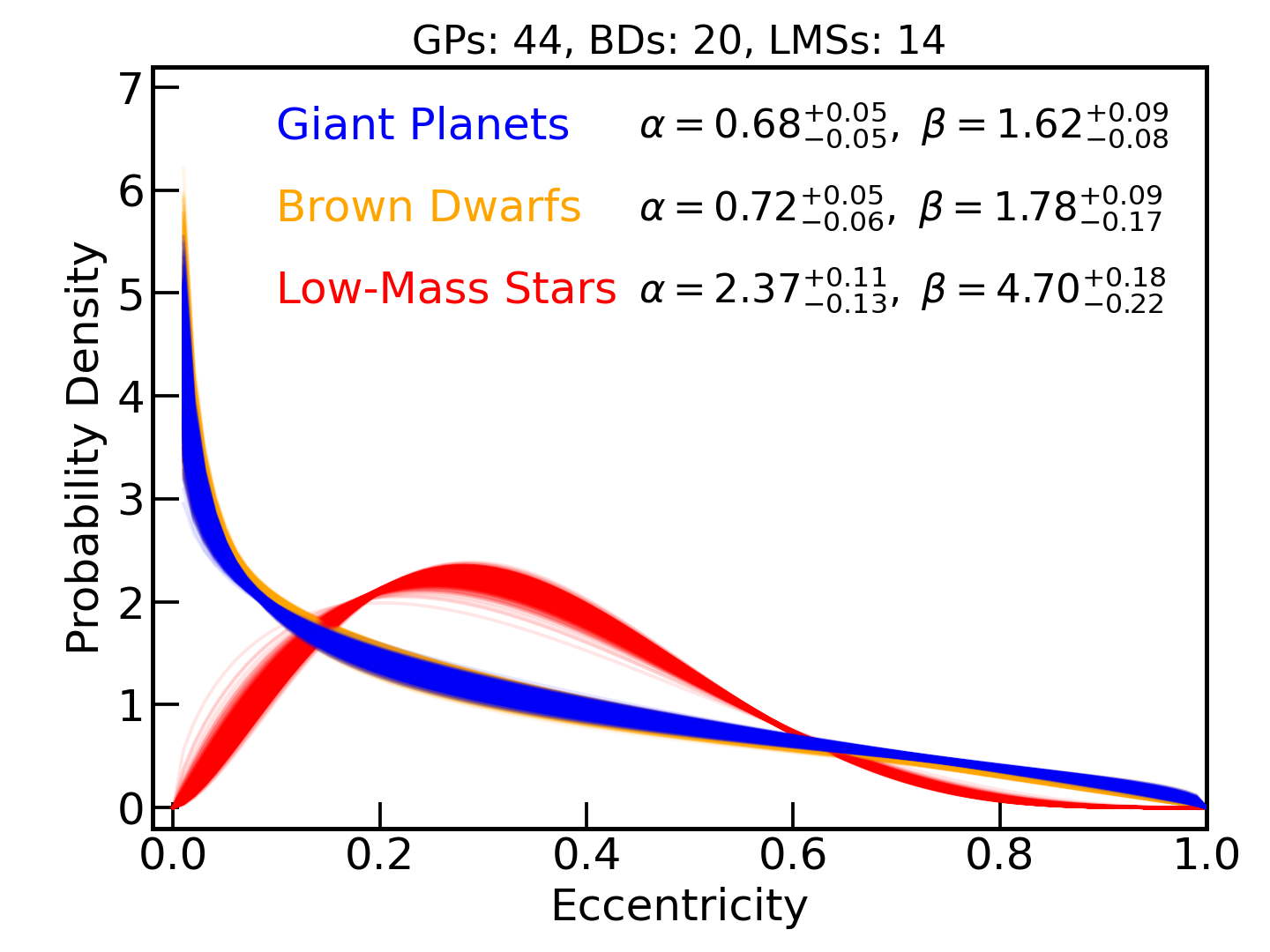}}

\caption{Same as Figure~\ref{betafit} but using $a/R_\ast$ to select long-period systems. From top to bottom, they are results using $a/R_\ast\geq10$, $a/R_\ast\geq15$ and $a/R_\ast\geq20$. The number of companions in each group is shown in the title of each panel.}
\label{betafit_a_Rs_cut}
\end{figure}

Since a more fundamental parameter related to the tidal circularization timescale is scaled semi-major axis, we repeat our analysis by selecting transiting long-period companions with $a/R_\ast$ instead. We investigate three different cuts: $a/R_\ast\geq10$, $a/R_\ast\geq15$ and $a/R_\ast\geq20$, and we perform the same Beta distribution fit as described above. The results are presented in Figure~\ref{betafit_a_Rs_cut}. We find that transiting long-period giant planets and brown dwarfs always prefer low eccentricities in all cases with similar eccentricity distributions, agreeing with our conclusion using the orbital period cut. The low-mass stars, however, only show an eccentricity peak at about 0.3 when we use the $a/R_\ast\geq15$ and $a/R_\ast\geq20$ cuts. We suspect that tidal effect may play a role in circularizing the orbits of some companions with $10 < a/R_\ast<15$.

\section{Discussions}\label{Discussions}

\subsection{Comparison to Wide-Orbit Cold Worlds}

Based on a sample of transiting long-period giant planets, brown dwarfs and low-mass stars with $10 \leq P\lesssim 1000$~days, we find that 1) the eccentricity distribution of giant planets resembles that of brown dwarfs; 2) the eccentricity distribution of low-mass stars is distinct from the other two populations with a peak at about 0.3. Using scaled semi-major axis ($a/R_\ast\geq 10,15,20$) instead to select the transiting long-period samples, we find that the conclusions do not change except that the peak at about 0.3 in the eccentricity distribution of low-mass stars disappears when we apply the $a/R_\ast\geq 10$ cut (Figure~\ref{betafit_a_Rs_cut}), which we attribute to tidal circularization. The semi-major axis range of our sample is roughly 0.1-1.5 AU. Therefore, our results complement the previous studies from \cite{Bowler2020}, \cite{Nagpal2023} and \cite{Doo2023}, who investigate the eccentricities of cold giants and brown dwarf companions detected by direct imaging. All works reported that the underlying eccentricity distributions of imaged cold gas giants and brown dwarfs within the semi-major axis range 5-100 AU are significantly different: cold brown dwarfs have a broad eccentricity distribution with a peak locating at 0.6–0.9 while imaged gas giants prefer a lower eccentricity around 0.05–0.25 \citep[see Figure~19 and 20 in][]{Bowler2020}. By performing a similar Beta distribution analysis, \cite{Bowler2020} found that the best-fit shape parameters of cold giant planets and brown dwarfs are [$\alpha \approx 30$, $\beta \approx 200$] and [$\alpha \approx 2.3$, $\beta \approx 1.65$], respectively. Interestingly, the eccentricity distribution of inner giant planets and brown dwarfs studied in this work are notably different from outer cold ones. We put forward two possible hypotheses to explain such a discrepancy. 

The first scenario is straightforward: warm Jupiters and brown dwarfs went through a similar evolution history.
Under this framework, both giant planets and brown dwarfs initially form far away from the hosts in two different channels, as suggested by \cite{Bowler2020}, resulting in different eccentricity distributions. Specifically, giant planets might generate from core accretion, which tends to induce low orbital eccentricities while brown dwarfs predominantly form by gravitational instability in massive disks that could inherit higher eccentricities. Subsequently, cold giant planets and brown dwarfs experience similar dynamic evolutions such as companion-disk interaction and companion-companion scattering \citep{Alqasim2025}, leading to analogous eccentricity distributions we finally saw in Figure~\ref{betafit}. 

The second hypothesis is that the eccentricity distributions of inner gas giants and brown dwarfs are jointly shaped by two different sub-groups. Similar as the previous scenario, the giant planets and brown dwarfs were born at exterior orbits with different eccentricity distributions. During the migration to inner orbits with $10\leq P\lesssim 1000$ days, some giant planets may remain dynamically stable all the time and preserve small eccentricities \citep{Teyssandier2020} while a part of them undergo orbital instability through scattering, leaving diverse eccentricities \citep{Ida2013}. The superposition of these two sub-groups yields the final eccentricity distribution seen in Figure~\ref{betafit}. Regarding the cold giant planets detected via direct imaging, they seem to prefer low-eccentricity orbits \citep{Bowler2020} because they probably do not encounter dynamically hot interactions after formation. In terms of brown dwarfs, considering that they originally formed through disk instability in the outer regions of massive disks, they tend to have high eccentricities at the beginning with a tentative peak at about 0.6–0.9, a feature seen in the direct imaging results \citep{Bowler2020}. However, their semi-major axes and eccentricities might be damped through dynamical friction with the disk since the formed brown dwarfs are sufficiently less massive than the disk \citep{Ida2020}. After migrating inwards to orbits with $10\leq P\lesssim 1000$ days, some brown dwarfs may still keep the high eccentricities inherited during formation while the others have damped low eccentricities. A combination of these two brown dwarf subgroups ultimately produces the eccentricity distribution in Figure~\ref{betafit}. Under this framework, the eccentricities of long-period giant planets and brown dwarfs with $10\leq P\lesssim 1000$ days have similar distributions by coincident since they are actually shaped by different formation and evolution mechanisms.

Finally, it should also be mentioned that the low-mass stars in our sample have a peak eccentricity around 0.3. Such a feature has also been observed in radial velocity surveys decades ago \citep[e.g.,][]{Duquennoy1991} and recent astrometry studies \citep[e.g.,][]{Wu2025} for more massive wide binaries. The results hint that the evolution of low-mass stars is probably similar to massive stellar binary systems. However, the origin of the peak at about 0.3 in the eccentricity distribution is still unclear. Theoretical works suggest that the orbits of binary systems could be excited by circumbinary disks \citep{Tokovinin2020,Lai2023,Siwek2023} and reach a high eccentricity at about 0.3, depending on intrinsic properties of the system such as mass ratio \citep{Valli2024}. In the meantime, multi components like outer massive companion \citep{Eggleton2006,Fabrycky2007,Naoz2014} as well as flyby encounters \citep{Heggie1996} should not be neglected, both of which could excite orbits. More recently, \cite{Wu2025} proposed that the eccentricity peak of close stellar binaries at about 0.3 could also be due to the ejection of brown dwarf objects. 

\subsection{Future Prospects}

The number of transiting long-period brown dwarfs and low-mass stars, particularly those have periods $P>100$~days with true mass measured (see Figure~\ref{evolution_track}), are still limited. Despite that, the Gaia DR4 and DR5 shall remedy the situation. Through the astrometry method, one could determine the true companion mass and the orbital eccentricity simultaneously \citep[e.g.,][]{Gan2023gaia,Stefansson2025}. In fact, numerous of astrometric binary stars and sub-stellar objects have already been released \citep{Halbwachs2023,Holl2023}. A larger sample will enable further exploration of the eccentricity trends of different stellar types, metallicities, periods, companion masses, etc, allowing for a more comprehensive mapping of eccentricity distribution and a more detailed comparison with predictions from different theories. 

Additionally, different dynamical origins of brown dwarfs are supposed to leave specific signatures on the stellar obliquities. Dynamically violent interactions such as high-eccentricity migration may cause misalignment between the orbital angular momentum axis of the companion and the spin axis of the host star. Instead, we expect to observe well-aligned systems with low obliquities if they move inwards via quiescent processes like disk migration. To date, only eight brown dwarf systems have their obliquity measured \citep{Triaud2009,Siverd2012,Triaud2013,Zhou2019hatp70,Ferreira2024,Giacalone2024,Brady2025,Doyle2025}, and all of them are short-period ($P<10$~days) systems and are consistent with aligned orbits within about $1\sigma$, hinting a dynamically quiet history. Consequently, it is crucial to enlarge this small sample and extend it to long-period brown dwarfs to further understand the evolution of such systems.

\section{Conclusions}\label{Conclusions}

In this paper, we report the discovery of \tar b, a long-period brown dwarf transiting a low-mass M5V star confirmed by space and ground photometric, spectroscopic and imaging observations. The brown dwarf companion has a mass of $72.4\pm 4.1\,{\rm M_J}$ and a radius of $0.84\pm0.07\,{\rm R_J}$ with a period of about 32 days on an eccentric orbit ($e=0.187\pm0.002$). Putting \tar b into a broader context, we find that the majority of eccentric brown dwarfs and low-mass stars with $e>0.3$ have mass between 42.5 and 110\,$\rm M_J$ with orbital periods greater than 10 days.

We further compare the eccentricities of transiting long-period giant planets ($0.3 \leq M_c<13.6\,{\rm M_J}$), brown dwarfs ($13.6 \leq M_c<80\,{\rm M_J}$) and low-mass stars ($80 \leq M_c\leq 150\,{\rm M_J}$) with orbital periods $10\leq P\lesssim 1000$~days (about 0.1-1.5 AU). We summarize our findings below.

\begin{enumerate}[(i)]
\item The eccentricity distributions of transiting long-period giant planets and brown dwarfs are almost identical ($\alpha = 0.64^{+0.07}_{-0.06}$ and $\beta=1.55^{+0.12}_{-0.10}$ for giant planets, $\alpha=0.78^{+0.08}_{-0.12}$ and $\beta=1.80^{+0.15}_{-0.20}$ for brown dwarfs): both favoring low-eccentricity orbits with a long tail toward high eccentricities.

\item Inner giant planets and brown dwarfs we studied in this work have eccentricity behaviors significantly different from outer cold counterparts within 5-100 AU detected by the direct imaging method ($\alpha \approx 30$ and $\beta \approx 200$ for imaged giant planets, $\alpha \approx 2.30$ and $\beta \approx 1.65$ for imaged brown dwarfs), where brown dwarfs show a broad eccentricity distribution with a peak locating at 0.6-0.9 but giant planets display a preference for low eccentricities between 0.05 and 0.25 \citep{Bowler2020}. 

\item Unlike the giant planet and brown dwarf population, transiting low-mass stars show a peak eccentricity at about 0.3 ($\alpha=2.42^{+0.12}_{-0.13}$ and \hbox{$\beta=5.12^{+0.20}_{-0.22}$}), similar to more massive wide binary systems. Low-mass stars with circular orbits or highly eccentric orbits ($e>0.6$) are rarer compared with the giant planets and brown dwarfs.

\end{enumerate}

The similarity in the eccentricity distributions of transiting long-period giant planets and brown dwarfs might be due to a similar dynamical evolution history or both of them contain two sub-groups that jointly shape similar eccentricity distributions. A larger sample size and follow-up obliquity studies will enable more comprehensive insights. 

\section{Acknowledgments}
We thank the anonymous referee for the constructive comments that improve the quality of this work. We also thank Noah Vowell and Khalid Barkaoui for kindly sharing their brown dwarf and low-mass star catalogs, without which this work would not be possible. We are grateful to Doug Lin, Yanqin Wu and Dong Lai for the useful discussions on the eccentricities of brown dwarfs and binary systems as well as dynamical processes. 

T.G. and S.M. acknowledge support by the National Natural Science Foundation of China (Grant No. 12133005). The work of I.A.S. was conducted under the state assignment of Lomonosov Moscow State University. Z.L.D. would like to thank the generous support of the MIT Presidential Fellowship, the MIT Collamore-Rogers Fellowship, and the MIT Teaching Development Fellowship and to acknowledge that this material is based upon work supported by the National Science Foundation Graduate Research Fellowship under Grant No. 1745302.

This research uses data obtained through the Telescope Access Program (TAP), which is funded by the National Astronomical Observatories, Chinese Academy of Sciences, and the Special Fund for Astronomy from the Ministry of Finance. This work is partly supported by the Natural Science and Engineering Research Council of Canada, the Canadian Space Agency and the Trottier Family Foundation through their support of the Trottier Institute for Research on Exoplanets (IREx). This work benefited from support of the Fonds de recherche du Québec – Nature et technologies (FRQNT), through the Center for Research in Astrophysics of Quebec. 

Based on observations obtained at the Canada-France-Hawaii Telescope (CFHT) which is operated from the summit of Maunakea by the National Research Council of Canada, the Institut National des Sciences de l'Univers of the Centre National de la Recherche Scientifique of France, and the University of Hawaii. The observations at the Canada-France-Hawaii Telescope were performed with care and respect from the summit of Maunakea which is a significant cultural and historic site. Based on observations obtained with SPIRou, an international project led by Institut de Recherche en Astrophysique et Planétologie, Toulouse, France.

Some of the observations in this paper made use of the High-Resolution Imaging instrument ‘Alopeke and were obtained under Gemini LLP Proposal Number: GN-2025A-DD-101. ‘Alopeke was funded by the NASA Exoplanet Exploration Program and built at the NASA Ames Research Center by Steve B. Howell, Nic Scott, Elliott P. Horch, and Emmett Quigley. Alopeke was mounted on the Gemini North telescope of the international Gemini Observatory, a program of NSF’s OIR Lab, which is managed by the Association of Universities for Research in Astronomy (AURA) under a cooperative agreement with the National Science Foundation. on behalf of the Gemini partnership: the National Science Foundation (United States), National Research Council (Canada), Agencia Nacional de Investigación y Desarrollo (Chile), Ministerio de Ciencia, Tecnología e Innovación (Argentina), Ministério da Ciência, Tecnologia, Inovações e Comunicações (Brazil), and Korea Astronomy and Space Science Institute (Republic of Korea).

Funding for the TESS mission is provided by NASA's Science Mission Directorate. We acknowledge the use of public TESS data from pipelines at the TESS Science Office and at the TESS Science Processing Operations Center. Resources supporting this work were provided by the NASA High-End Computing (HEC) Program through the NASA Advanced Supercomputing (NAS) Division at Ames Research Center for the production of the SPOC data products. This research has made use of the Exoplanet Follow-up Observation Program website, which is operated by the California Institute of Technology, under contract with the National Aeronautics and Space Administration under the Exoplanet Exploration Program. This paper includes data collected by the \tess\ mission, which are publicly available from the Mikulski Archive for Space Telescopes\ (MAST).


\facilities{TESS, CFHT/SPIRou, Keck II/NIRES, Gemini-North/‘Alopeke, SAI-2.5m, ASP-0.36m, Gaia, ZTF}

\software{lightkurve \citep{lightkurvecollaboration}, AstroImageJ \citep{Collins2017}, juliet \citep{juliet}, batman \citep{Kreidberg2015}, radvel \citep{Fulton2018}, Scipy \citep{Virtanen2020}, corner \citep{Foreman2016}}



\appendix

\restartappendixnumbering

\section{Normalized TESS light curves}

Figure~\ref{tesslc} shows the normalized and detrended TESS light curves of \tar. 

\renewcommand{\theHfigure}{A.\arabic{figure}} 
\setcounter{figure}{0}

\begin{figure*}
\centering
\includegraphics[width=0.99\textwidth]{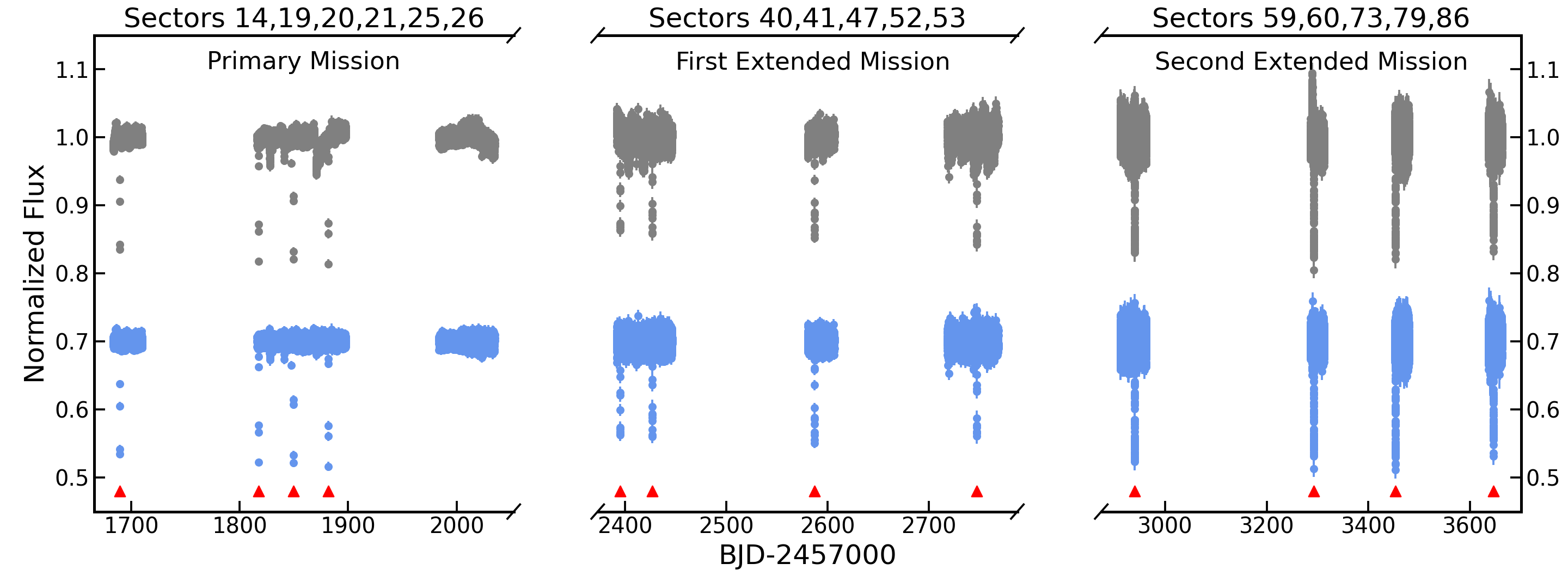}
\caption{The normalized TESS light curves of \tar. The left, middle and right panels are data from Primary, First and Second Extended Missions. The light curve from the Primary Mission was extracted with a $3\times 3$ pixel box aperture while the light curves from the First and Second Extended Missions came from QLP and SPOC pipelines. The different transit depths seen in three panels are likely due to the different aperture sizes, resulting in different light dilutions. The raw light curves are shown above in gray while the detrended light curves are presented below in blue. The transits of \tar b are marked with red triangles. The transit event in Sectors 25 and 26 happened during the data gaps so it does not show up.}
\label{tesslc}
\end{figure*}

\section{SPIRou Radial Velocities}

\renewcommand{\theHtable}{B.\arabic{table}}
\setcounter{table}{0}

All SPIRou radial velocities we obtained for \tar\ are listed in Table~\ref{RVtable}.

\begin{table}[h]
    \caption{Radial velocities measurements for \tar\ from SPIRou.}
    \begin{tabular}{ccc}
        \hline\hline
        BJD       &RV (m~s$^{-1}$) &$\sigma_{\rm RV}$ (m~s$^{-1}$)      \\\hline
        2460065.995	&-42867.1 	&16.8 \\
2460066.009	&-42881.0 	&17.5 \\
2460067.997	&-46980.4 	&25.2 \\
2460068.012	&-46946.8 	&39.8 \\
2460072.960	&-58244.1 	&16.7 \\
2460072.974	&-58238.5 	&17.0 \\
2460074.957	&-58835.2 	&16.8 \\
2460074.972	&-58773.4 	&17.1 \\
2460076.988	&-56064.7 	&21.6 \\
2460077.002	&-56009.5 	&21.7 \\
2460096.888	&-40897.2 	&26.0 \\
2460096.902	&-40904.2 	&29.5 \\
2460098.911	&-44622.0 	&22.2 \\
2460098.926	&-44586.1 	&22.1 \\
2460101.898	&-51580.8 	&22.8 \\
2460101.912	&-51605.9 	&24.7 \\
2460103.887	&-56373.3 	&15.2 \\
2460103.901	&-56369.3 	&15.2 \\
2460123.826	&-36998.1 	&17.7 \\
2460123.840	&-37004.1 	&17.2 \\
2460124.871	&-37266.9 	&16.9 \\
2460124.885	&-37248.2 	&16.6 \\
2460125.861	&-37773.0 	&19.4 \\
2460125.875	&-37772.1 	&20.4 \\
         \hline
    \end{tabular}
    \label{RVtable}
\end{table}

\section{The Keck/NIRES spectrum and SED of \tar}

\renewcommand{\theHfigure}{C.\arabic{figure}} 
\setcounter{figure}{0}

\begin{figure*}
\centering
\includegraphics[width=0.57\textwidth]{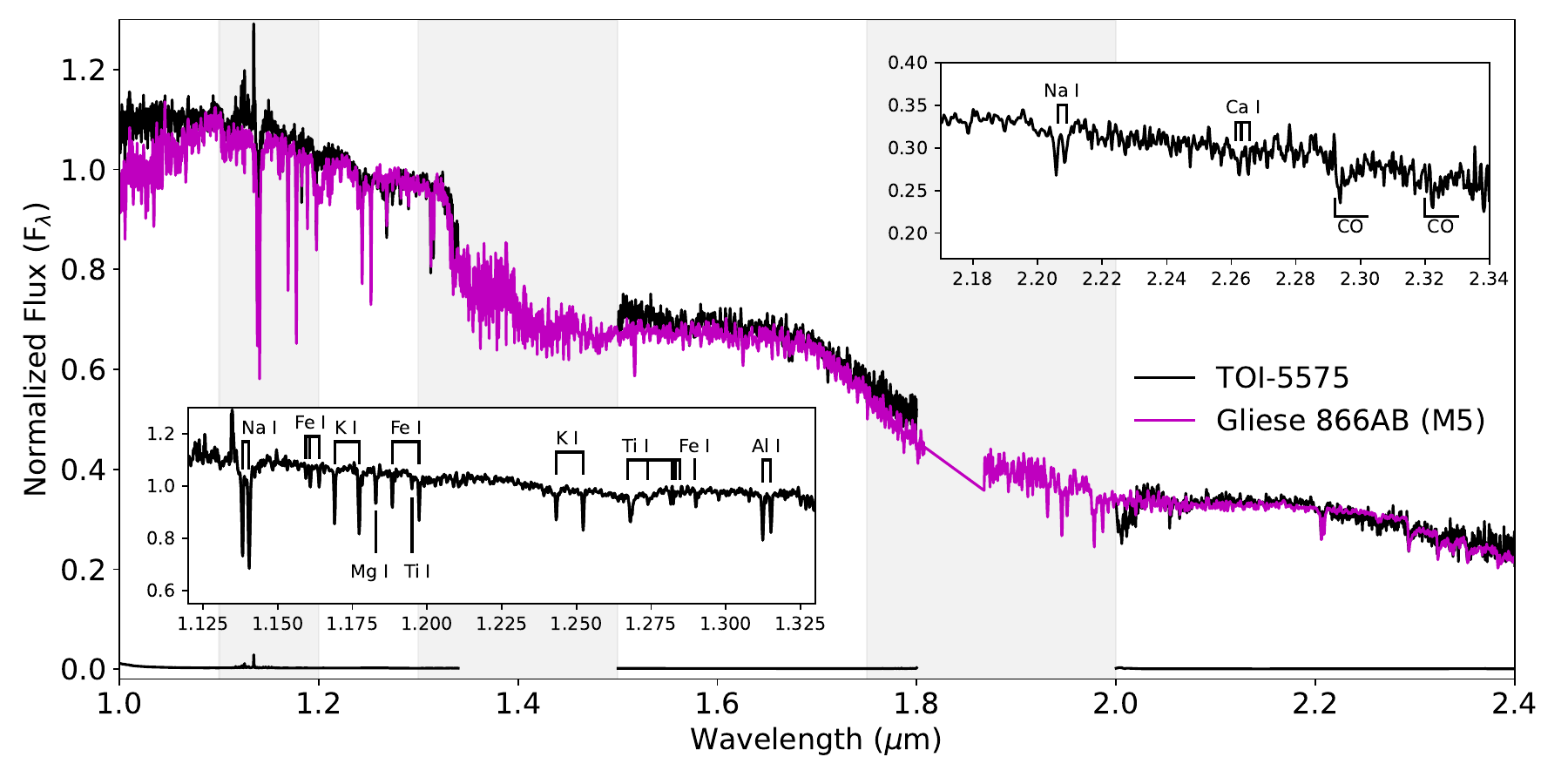}
\includegraphics[width=0.42\textwidth]{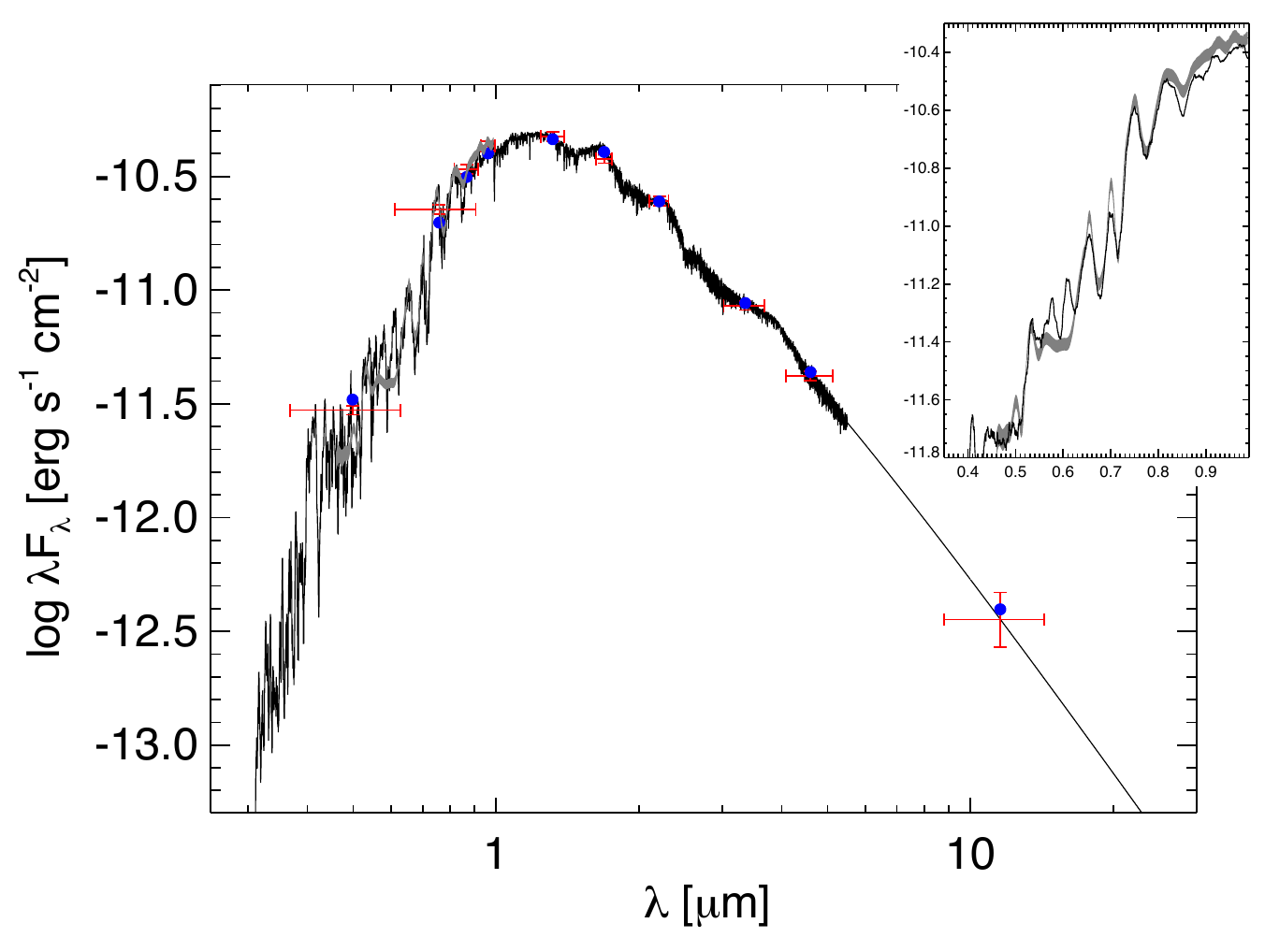}
\caption{{\it Left panel:} Keck/NIRES spectrum of TOI-5575 (black line) compared to the equivalent data for the M5 standard Gliese 866AB (magenta line; \citealt{2005ApJ...623.1115C}). Both spectra are normalized at 1.3~$\mu$m. Locations of major H$_2$O and CO molecular absorption bands are labeled, as are regions of strong telluric absorption (grey boxes, masked in spectrum of TOI-5575). The inset boxes show close-ups of the 1.12--1.33~$\mu$m and 2.18--2.34~$\mu$m regions with primary atomic and molecular features labeled. {\it Right panel:} The SED of \tar. The red crosses are the observed photometric
measurements, where the horizontal bars represent the effective width of the passband. All blue dots represent the model fluxes from the best-fit PHOENIX atmosphere model (black solid lines). The Gaia spectrum is shown as a gray swathe on the inset top right panel.} 
\label{fig:sed}
\end{figure*}

Figure~\ref{fig:sed} shows the Keck/NIRES spectrum as well as broadband photometry of \tar\ along with the best-fit model. 

\section{Speckle imaging results}

\renewcommand{\theHfigure}{D.\arabic{figure}} 
\setcounter{figure}{0}

\begin{figure}
\centering
\includegraphics[width=0.49\textwidth]{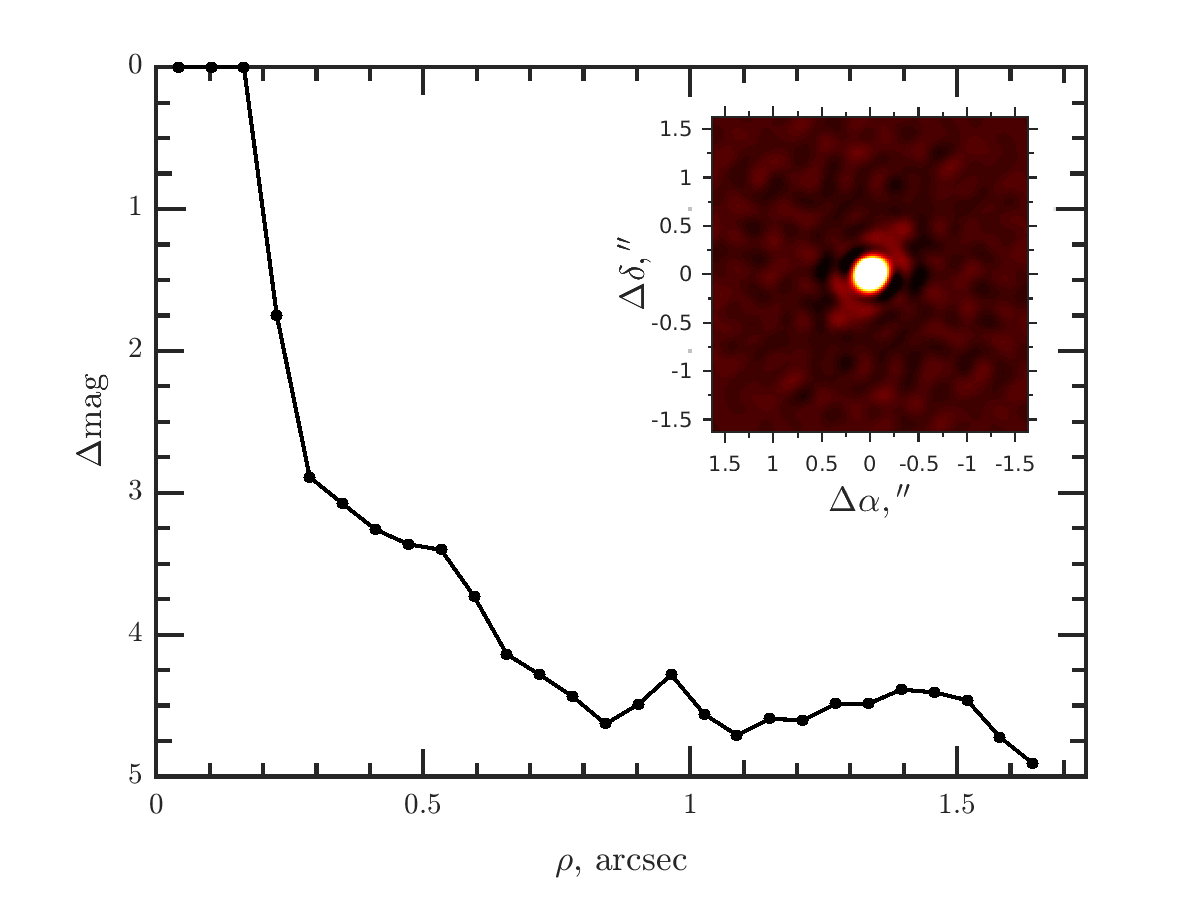}
\includegraphics[width=0.49\textwidth]{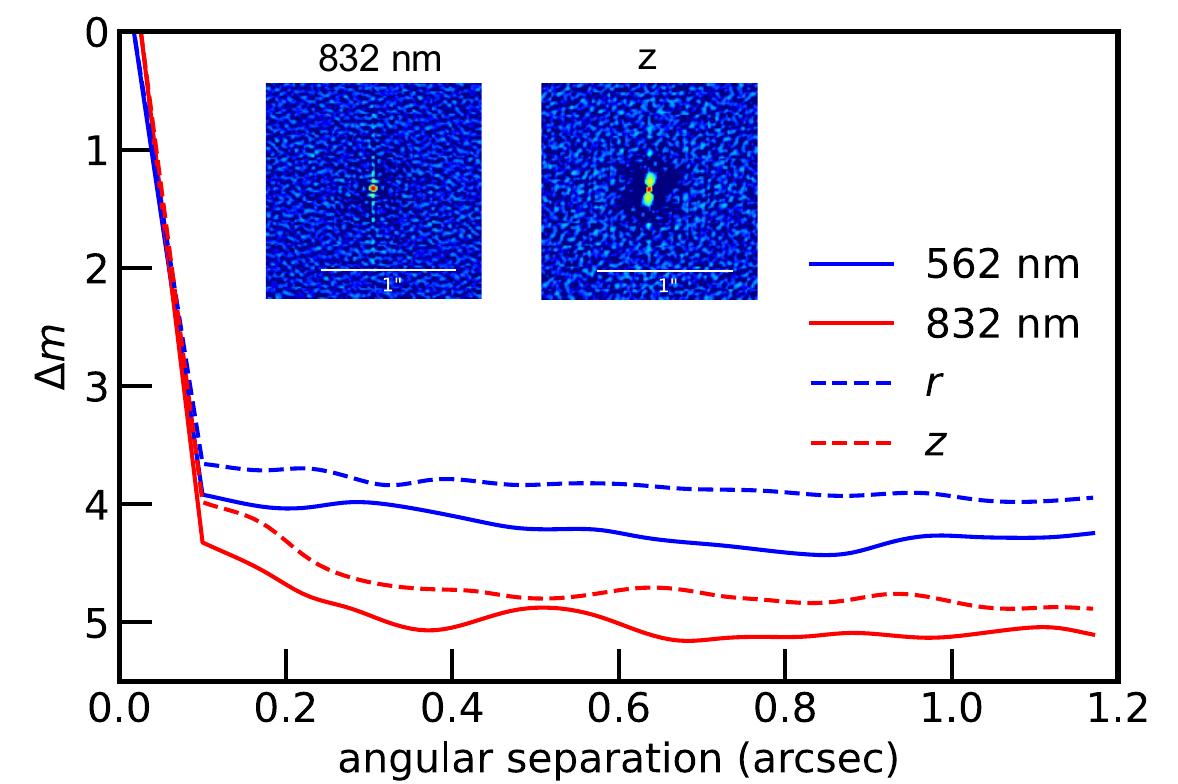}
\caption{{\it Top panel:} Secondary component detection limits obtained using speckle imaging obtained with speckle polarimeter on the SAI-2.5m telescope. {\it Bottom panel:} Gemini high-resolution speckle imaging $5\sigma$ contrast limits as a function of angular separation. There are hints of a close companion in the reconstructed images, but these observations were taken at high airmass, and the orientation is consistent with the star being ``smeared'' along the parallactic angle, rather than being a close-in binary.}
\label{fig:spp}
\end{figure}

Figure~\ref{fig:spp} shows the speckle imaging results from SAI-2.5m and Gemini. 

\bibliography{planet}{}

\begin{thebibliography}{}
\expandafter\ifx\csname natexlab\endcsname\relax\def\natexlab#1{#1}\fi
\providecommand{\url}[1]{\href{#1}{#1}}
\providecommand{\dodoi}[1]{doi:~\href{http://doi.org/#1}{\nolinkurl{#1}}}
\providecommand{\doeprint}[1]{\href{http://ascl.net/#1}{\nolinkurl{http://ascl.net/#1}}}
\providecommand{\doarXiv}[1]{\href{https://arxiv.org/abs/#1}{\nolinkurl{https://arxiv.org/abs/#1}}}

\bibitem[{{Akeson} {et~al.}(2013){Akeson}, {Chen}, {Ciardi}, {Crane}, {Good}, {Harbut}, {Jackson}, {Kane}, {Laity}, {Leifer}, {Lynn}, {McElroy}, {Papin}, {Plavchan}, {Ram{\'{\i}}rez}, {Rey}, {von Braun}, {Wittman}, {Abajian}, {Ali}, {Beichman}, {Beekley}, {Berriman}, {Berukoff}, {Bryden}, {Chan}, {Groom}, {Lau}, {Payne}, {Regelson}, {Saucedo}, {Schmitz}, {Stauffer}, {Wyatt}, \& {Zhang}}]{Akeson2013}
{Akeson}, R.~L., {Chen}, X., {Ciardi}, D., {et~al.} 2013, \pasp, 125, 989, \dodoi{10.1086/672273}

\bibitem[{{Alqasim} {et~al.}(2025){Alqasim}, {Hirano}, {Hori}, {Kawata}, {Livingston}, {Howell}, {Lanza}, {Mann}, {Ziegler}, {Brice{\~n}o}, {Beichman}, {Ciardi}, {Strakhov}, {Lund}, \& {Law}}]{Alqasim2025}
{Alqasim}, A., {Hirano}, T., {Hori}, Y., {et~al.} 2025, \mnras, \dodoi{10.1093/mnras/staf418}

\bibitem[{{Artigau} {et~al.}(2021){Artigau}, {H{\'e}brard}, {Cadieux}, {Vandal}, {Cook}, {Doyon}, {Gagn{\'e}}, {Moutou}, {Martioli}, {Frasca}, {Jahandar}, {Lafreni{\`e}re}, {Malo}, {Donati}, {Cort{\'e}s-Zuleta}, {Boisse}, {Delfosse}, {Carmona}, {Fouqu{\'e}}, {Morin}, {Rowe}, {Marino}, {Papini}, {Ciardi}, {Lund}, {Martins}, {Pelletier}, {Arnold}, {Bouchy}, {Forveille}, {Santos}, {Bonfils}, {Figueira}, {Fausnaugh}, {Ricker}, {Latham}, {Seager}, {Winn}, {Jenkins}, {Ting}, {Torres}, \& {Gomes da Silva}}]{Artigau2021}
{Artigau}, {\'E}., {H{\'e}brard}, G., {Cadieux}, C., {et~al.} 2021, \aj, 162, 144, \dodoi{10.3847/1538-3881/ac096d}

\bibitem[{{Artigau} {et~al.}(2022){Artigau}, {Cadieux}, {Cook}, {Doyon}, {Vandal}, {Donati}, {Moutou}, {Delfosse}, {Fouqu{\'e}}, {Martioli}, {Bouchy}, {Parsons}, {Carmona}, {Dumusque}, {Astudillo-Defru}, {Bonfils}, \& {Mignon}}]{Artigau2022}
{Artigau}, {\'E}., {Cadieux}, C., {Cook}, N.~J., {et~al.} 2022, \aj, 164, 84, \dodoi{10.3847/1538-3881/ac7ce6}

\bibitem[{{Baraffe} {et~al.}(2003){Baraffe}, {Chabrier}, {Barman}, {Allard}, \& {Hauschildt}}]{Baraffe2003}
{Baraffe}, I., {Chabrier}, G., {Barman}, T.~S., {Allard}, F., \& {Hauschildt}, P.~H. 2003, \aap, 402, 701, \dodoi{10.1051/0004-6361:20030252}

\bibitem[{{Barkaoui} {et~al.}(2025){Barkaoui}, {Sebastian}, {Z{\'u}{\~n}iga-Fern{\'a}ndez}, {Triaud}, {Rackham}, {Burgasser}, {Carmichael}, {Gillon}, {Theissen}, {Softich}, {Rojas-Ayala}, {Srdoc}, {Soubkiou}, {Fukui}, {Timmermans}, {Stalport}, {Burdanov}, {Ciardi}, {Collins}, {Davis}, {Davoudi}, {de Wit}, {Demory}, {Deveny}, {Dransfield}, {Ducrot}, {Florian}, {Gan}, {G{\'o}mez Maqueo Chew}, {Hooton}, {Howell}, {Jenkins}, {Littlefield}, {Mart{\'\i}n}, {Murgas}, {Niraula}, {Palle}, {Pedersen}, {Pozuelos}, {Queloz}, {Ricker}, {Schwarz}, {Seager}, {Shporer}, {Scott}, {Stockdale}, \& {Winn}}]{Barkaoui2025}
{Barkaoui}, K., {Sebastian}, D., {Z{\'u}{\~n}iga-Fern{\'a}ndez}, S., {et~al.} 2025, \aap, 696, A44, \dodoi{10.1051/0004-6361/202453508}

\bibitem[{{Beaug{\'e}} \& {Nesvorn{\'y}}(2012)}]{Beauge2012}
{Beaug{\'e}}, C., \& {Nesvorn{\'y}}, D. 2012, \apj, 751, 119, \dodoi{10.1088/0004-637X/751/2/119}

\bibitem[{{Bellm} {et~al.}(2019){Bellm}, {Kulkarni}, {Graham}, {Dekany}, {Smith}, {Riddle}, {Masci}, {Helou}, {Prince}, {Adams}, {Barbarino}, {Barlow}, {Bauer}, {Beck}, {Belicki}, {Biswas}, {Blagorodnova}, {Bodewits}, {Bolin}, {Brinnel}, {Brooke}, {Bue}, {Bulla}, {Burruss}, {Cenko}, {Chang}, {Connolly}, {Coughlin}, {Cromer}, {Cunningham}, {De}, {Delacroix}, {Desai}, {Duev}, {Eadie}, {Farnham}, {Feeney}, {Feindt}, {Flynn}, {Franckowiak}, {Frederick}, {Fremling}, {Gal-Yam}, {Gezari}, {Giomi}, {Goldstein}, {Golkhou}, {Goobar}, {Groom}, {Hacopians}, {Hale}, {Henning}, {Ho}, {Hover}, {Howell}, {Hung}, {Huppenkothen}, {Imel}, {Ip}, {Ivezi{\'c}}, {Jackson}, {Jones}, {Juric}, {Kasliwal}, {Kaspi}, {Kaye}, {Kelley}, {Kowalski}, {Kramer}, {Kupfer}, {Landry}, {Laher}, {Lee}, {Lin}, {Lin}, {Lunnan}, {Giomi}, {Mahabal}, {Mao}, {Miller}, {Monkewitz}, {Murphy}, {Ngeow}, {Nordin}, {Nugent}, {Ofek}, {Patterson}, {Penprase}, {Porter}, {Rauch}, {Rebbapragada}, {Reiley}, {Rigault}, {Rodriguez}, {van Roestel}, {Rusholme}, {van
  Santen}, {Schulze}, {Shupe}, {Singer}, {Soumagnac}, {Stein}, {Surace}, {Sollerman}, {Szkody}, {Taddia}, {Terek}, {Van Sistine}, {van Velzen}, {Vestrand}, {Walters}, {Ward}, {Ye}, {Yu}, {Yan}, \& {Zolkower}}]{Bellm2019}
{Bellm}, E.~C., {Kulkarni}, S.~R., {Graham}, M.~J., {et~al.} 2019, \pasp, 131, 018002, \dodoi{10.1088/1538-3873/aaecbe}

\bibitem[{{Benedict} {et~al.}(2016){Benedict}, {Henry}, {Franz}, {McArthur}, {Wasserman}, {Jao}, {Cargile}, {Dieterich}, {Bradley}, {Nelan}, \& {Whipple}}]{Benedict2016}
{Benedict}, G.~F., {Henry}, T.~J., {Franz}, O.~G., {et~al.} 2016, \aj, 152, 141, \dodoi{10.3847/0004-6256/152/5/141}

\bibitem[{{Boss}(2000)}]{Boss2000}
{Boss}, A.~P. 2000, \apjl, 536, L101, \dodoi{10.1086/312737}

\bibitem[{{Bowler} {et~al.}(2020){Bowler}, {Blunt}, \& {Nielsen}}]{Bowler2020}
{Bowler}, B.~P., {Blunt}, S.~C., \& {Nielsen}, E.~L. 2020, \aj, 159, 63, \dodoi{10.3847/1538-3881/ab5b11}

\bibitem[{{Brady} {et~al.}(2025){Brady}, {Bean}, {Stef{\'a}nsson}, {Brown}, {Seifahrt}, {Basant}, {Das}, {Luque}, \& {St{\"u}rmer}}]{Brady2025}
{Brady}, M., {Bean}, J.~L., {Stef{\'a}nsson}, G., {et~al.} 2025, \aj, 169, 64, \dodoi{10.3847/1538-3881/ad9c66}

\bibitem[{{Burgasser} {et~al.}(2003){Burgasser}, {Kirkpatrick}, {Reid}, {Brown}, {Miskey}, \& {Gizis}}]{Burgasser2003}
{Burgasser}, A.~J., {Kirkpatrick}, J.~D., {Reid}, I.~N., {et~al.} 2003, \apj, 586, 512, \dodoi{10.1086/346263}

\bibitem[{{Carmichael} {et~al.}(2022){Carmichael}, {Irwin}, {Murgas}, {Pall{\'e}}, {Stassun}, {Bartnik}, {Collins}, {de Leon}, {Esparza-Borges}, {Fedewa}, {Fong}, {Fukui}, {Jenkins}, {Kagetani}, {Latham}, {Lund}, {Mann}, {Moldovan}, {Morgan}, {Narita}, {Painter}, {Parviainen}, {Quintana}, {Ricker}, {Schulte}, {Schwarz}, {Seager}, {Sokolovsky}, {Twicken}, \& {Winn}}]{Carmichael2022}
{Carmichael}, T.~W., {Irwin}, J.~M., {Murgas}, F., {et~al.} 2022, \mnras, 514, 4944, \dodoi{10.1093/mnras/stac1666}

\bibitem[{{Carrera} {et~al.}(2019){Carrera}, {Raymond}, \& {Davies}}]{Carrera2019}
{Carrera}, D., {Raymond}, S.~N., \& {Davies}, M.~B. 2019, \aap, 629, L7, \dodoi{10.1051/0004-6361/201935744}

\bibitem[{{Charbonneau} {et~al.}(2005){Charbonneau}, {Allen}, {Megeath}, {Torres}, {Alonso}, {Brown}, {Gilliland}, {Latham}, {Mandushev}, {O'Donovan}, \& {Sozzetti}}]{Charbonneau2005}
{Charbonneau}, D., {Allen}, L.~E., {Megeath}, S.~T., {et~al.} 2005, \apj, 626, 523, \dodoi{10.1086/429991}

\bibitem[{{Clark} {et~al.}(2024){Clark}, {van Belle}, {Horch}, {Lund}, {Ciardi}, {von Braun}, {Winters}, {Everett}, {Hartman}, \& {Llama}}]{Clark2024AJ....167...56C}
{Clark}, C.~A., {van Belle}, G.~T., {Horch}, E.~P., {et~al.} 2024, \aj, 167, 56, \dodoi{10.3847/1538-3881/ad0bfd}

\bibitem[{{Collins} {et~al.}(2017){Collins}, {Kielkopf}, {Stassun}, \& {Hessman}}]{Collins2017}
{Collins}, K.~A., {Kielkopf}, J.~F., {Stassun}, K.~G., \& {Hessman}, F.~V. 2017, \aj, 153, 77, \dodoi{10.3847/1538-3881/153/2/77}

\bibitem[{{Cook} {et~al.}(2022){Cook}, {Artigau}, {Doyon}, {Hobson}, {Martioli}, {Bouchy}, {Moutou}, {Carmona}, {Usher}, {Fouqu{\'e}}, {Arnold}, {Delfosse}, {Boisse}, {Cadieux}, {Vandal}, {Donati}, \& {Desli{\`e}res}}]{Cook2022}
{Cook}, N.~J., {Artigau}, {\'E}., {Doyon}, R., {et~al.} 2022, \pasp, 134, 114509, \dodoi{10.1088/1538-3873/ac9e74}

\bibitem[{{Cushing} {et~al.}(2005){Cushing}, {Rayner}, \& {Vacca}}]{2005ApJ...623.1115C}
{Cushing}, M.~C., {Rayner}, J.~T., \& {Vacca}, W.~D. 2005, \apj, 623, 1115, \dodoi{10.1086/428040}

\bibitem[{{Cushing} {et~al.}(2004){Cushing}, {Vacca}, \& {Rayner}}]{2004PASP..116..362C}
{Cushing}, M.~C., {Vacca}, W.~D., \& {Rayner}, J.~T. 2004, \pasp, 116, 362, \dodoi{10.1086/382907}

\bibitem[{{Cutri} {et~al.}(2003){Cutri}, {Skrutskie}, {van Dyk}, {Beichman}, {Carpenter}, {Chester}, {Cambresy}, {Evans}, {Fowler}, {Gizis}, {Howard}, {Huchra}, {Jarrett}, {Kopan}, {Kirkpatrick}, {Light}, {Marsh}, {McCallon}, {Schneider}, {Stiening}, {Sykes}, {Weinberg}, {Wheaton}, {Wheelock}, \& {Zacarias}}]{Cutri2003}
{Cutri}, R.~M., {Skrutskie}, M.~F., {van Dyk}, S., {et~al.} 2003, {2MASS All Sky Catalog of point sources.}

\bibitem[{{Dawson} \& {Johnson}(2012)}]{Dawson2012}
{Dawson}, R.~I., \& {Johnson}, J.~A. 2012, \apj, 756, 122, \dodoi{10.1088/0004-637X/756/2/122}

\bibitem[{{Dawson} \& {Johnson}(2018)}]{Dawson2018}
---. 2018, \araa, 56, 175, \dodoi{10.1146/annurev-astro-081817-051853}

\bibitem[{{Dawson} \& {Murray-Clay}(2013)}]{Dawson2013}
{Dawson}, R.~I., \& {Murray-Clay}, R.~A. 2013, \apjl, 767, L24, \dodoi{10.1088/2041-8205/767/2/L24}

\bibitem[{{D{\'\i}az} {et~al.}(2013){D{\'\i}az}, {Damiani}, {Deleuil}, {Almenara}, {Moutou}, {Barros}, {Bonomo}, {Bouchy}, {Bruno}, {H{\'e}brard}, {Montagnier}, \& {Santerne}}]{Diaz2013}
{D{\'\i}az}, R.~F., {Damiani}, C., {Deleuil}, M., {et~al.} 2013, \aap, 551, L9, \dodoi{10.1051/0004-6361/201321124}

\bibitem[{{Do {\'O}} {et~al.}(2023){Do {\'O}}, {O'Neil}, {Konopacky}, {Do}, {Martinez}, {Ruffio}, \& {Ghez}}]{Doo2023}
{Do {\'O}}, C.~R., {O'Neil}, K.~K., {Konopacky}, Q.~M., {et~al.} 2023, \aj, 166, 48, \dodoi{10.3847/1538-3881/acdc9a}

\bibitem[{{Donati} {et~al.}(2020){Donati}, {Kouach}, {Moutou}, {Doyon}, {Delfosse}, {Artigau}, {Baratchart}, {Lacombe}, {Barrick}, {H{\'e}brard}, {Bouchy}, {Saddlemyer}, {Par{\`e}s}, {Rabou}, {Micheau}, {Dolon}, {Reshetov}, {Challita}, {Carmona}, {Striebig}, {Thibault}, {Martioli}, {Cook}, {Fouqu{\'e}}, {Vermeulen}, {Wang}, {Arnold}, {Pepe}, {Boisse}, {Figueira}, {Bouvier}, {Ray}, {Feugeade}, {Morin}, {Alencar}, {Hobson}, {Castilho}, {Udry}, {Santos}, {Hernandez}, {Benedict}, {Vall{\'e}e}, {Gallou}, {Dupieux}, {Larrieu}, {Perruchot}, {Sottile}, {Moreau}, {Usher}, {Baril}, {Wildi}, {Chazelas}, {Malo}, {Bonfils}, {Loop}, {Kerley}, {Wevers}, {Dunn}, {Pazder}, {Macdonald}, {Dubois}, {Carri{\'e}}, {Valentin}, {Henault}, {Yan}, \& {Steinmetz}}]{Donati2020}
{Donati}, J.~F., {Kouach}, D., {Moutou}, C., {et~al.} 2020, \mnras, 498, 5684, \dodoi{10.1093/mnras/staa2569}

\bibitem[{{Dong} {et~al.}(2021){Dong}, {Huang}, {Dawson}, {Foreman-Mackey}, {Collins}, {Quinn}, {Lissauer}, {Beatty}, {Quarles}, {Sha}, {Shporer}, {Guo}, {Kane}, {Abe}, {Barkaoui}, {Benkhaldoun}, {Brahm}, {Bouchy}, {Carmichael}, {Collins}, {Conti}, {Crouzet}, {Dransfield}, {Evans}, {Gan}, {Ghachoui}, {Gillon}, {Grieves}, {Guillot}, {Hellier}, {Jehin}, {Jensen}, {Jord{\'a}n}, {Kamler}, {Kielkopf}, {M{\'e}karnia}, {Nielsen}, {Pozuelos}, {Radford}, {Schmider}, {Schwarz}, {Stockdale}, {Tan}, {Timmermans}, {Triaud}, {Wang}, {Ricker}, {Vanderspek}, {Latham}, {Seager}, {Winn}, {Jenkins}, {Mireles}, {Yahalomi}, {Morgan}, {Vezie}, {Quintana}, {Rose}, {Smith}, \& {Shiao}}]{Dong2021}
{Dong}, J., {Huang}, C.~X., {Dawson}, R.~I., {et~al.} 2021, \apjs, 255, 6, \dodoi{10.3847/1538-4365/abf73c}

\bibitem[{{Doyle} {et~al.}(2025){Doyle}, {Ca{\~n}as}, {Libby-Roberts}, {Cegla}, {Stef{\'a}nsson}, {Anderson}, {Armstrong}, {Bender}, {Bayliss}, {Carmichael}, {Casewell}, {Kanodia}, {Lafarga}, {Lin}, {Mahadevan}, {Monson}, {Robertson}, \& {Veras}}]{Doyle2025}
{Doyle}, L., {Ca{\~n}as}, C.~I., {Libby-Roberts}, J.~E., {et~al.} 2025, \mnras, 536, 3745, \dodoi{10.1093/mnras/stae2819}

\bibitem[{{Duquennoy} \& {Mayor}(1991)}]{Duquennoy1991}
{Duquennoy}, A., \& {Mayor}, M. 1991, \aap, 248, 485

\bibitem[{{Eggleton} \& {Kisseleva-Eggleton}(2006)}]{Eggleton2006}
{Eggleton}, P.~P., \& {Kisseleva-Eggleton}, L. 2006, \apss, 304, 75, \dodoi{10.1007/s10509-006-9078-z}

\bibitem[{{El-Badry} {et~al.}(2023){El-Badry}, {Burdge}, {van Roestel}, \& {Rodriguez}}]{ElBadry2023}
{El-Badry}, K., {Burdge}, K.~B., {van Roestel}, J., \& {Rodriguez}, A.~C. 2023, The Open Journal of Astrophysics, 6, 33, \dodoi{10.21105/astro.2307.15729}

\bibitem[{{Espinoza}(2018)}]{Espinoza2018}
{Espinoza}, N. 2018, Research Notes of the American Astronomical Society, 2, 209, \dodoi{10.3847/2515-5172/aaef38}

\bibitem[{{Espinoza} {et~al.}(2019){Espinoza}, {Kossakowski}, \& {Brahm}}]{juliet}
{Espinoza}, N., {Kossakowski}, D., \& {Brahm}, R. 2019, \mnras, 490, 2262, \dodoi{10.1093/mnras/stz2688}

\bibitem[{{Fabrycky} \& {Tremaine}(2007)}]{Fabrycky2007}
{Fabrycky}, D., \& {Tremaine}, S. 2007, \apj, 669, 1298, \dodoi{10.1086/521702}

\bibitem[{{Ferreira dos Santos} {et~al.}(2024){Ferreira dos Santos}, {Rice}, {Wang}, \& {Wang}}]{Ferreira2024}
{Ferreira dos Santos}, T., {Rice}, M., {Wang}, X.-Y., \& {Wang}, S. 2024, \aj, 168, 145, \dodoi{10.3847/1538-3881/ad6b7f}

\bibitem[{{Foreman-Mackey}(2016)}]{Foreman2016}
{Foreman-Mackey}, D. 2016, The Journal of Open Source Software, 1, 24, \dodoi{10.21105/joss.00024}

\bibitem[{{Foreman-Mackey} {et~al.}(2017){Foreman-Mackey}, {Agol}, {Ambikasaran}, \& {Angus}}]{Foreman2017}
{Foreman-Mackey}, D., {Agol}, E., {Ambikasaran}, S., \& {Angus}, R. 2017, \aj, 154, 220, \dodoi{10.3847/1538-3881/aa9332}

\bibitem[{{Fulton} {et~al.}(2018){Fulton}, {Petigura}, {Blunt}, \& {Sinukoff}}]{Fulton2018}
{Fulton}, B.~J., {Petigura}, E.~A., {Blunt}, S., \& {Sinukoff}, E. 2018, \pasp, 130, 044504, \dodoi{10.1088/1538-3873/aaaaa8}

\bibitem[{{Gaia Collaboration} {et~al.}(2021){Gaia Collaboration}, {Brown}, {Vallenari}, {Prusti}, {de Bruijne}, {Babusiaux}, {Biermann}, {Creevey}, {Evans}, {Eyer}, {Hutton}, {Jansen}, {Jordi}, {Klioner}, {Lammers}, {Lindegren}, {Luri}, {Mignard}, {Panem}, {Pourbaix}, {Randich}, {Sartoretti}, {Soubiran}, {Walton}, {Arenou}, {Bailer-Jones}, {Bastian}, {Cropper}, {Drimmel}, {Katz}, {Lattanzi}, {van Leeuwen}, {Bakker}, {Cacciari}, {Casta{\~n}eda}, {De Angeli}, {Ducourant}, {Fabricius}, {Fouesneau}, {Fr{\'e}mat}, {Guerra}, {Guerrier}, {Guiraud}, {Jean-Antoine Piccolo}, {Masana}, {Messineo}, {Mowlavi}, {Nicolas}, {Nienartowicz}, {Pailler}, {Panuzzo}, {Riclet}, {Roux}, {Seabroke}, {Sordo}, {Tanga}, {Th{\'e}venin}, {Gracia-Abril}, {Portell}, {Teyssier}, {Altmann}, {Andrae}, {Bellas-Velidis}, {Benson}, {Berthier}, {Blomme}, {Brugaletta}, {Burgess}, {Busso}, {Carry}, {Cellino}, {Cheek}, {Clementini}, {Damerdji}, {Davidson}, {Delchambre}, {Dell'Oro}, {Fern{\'a}ndez-Hern{\'a}ndez}, {Galluccio}, {Garc{\'\i}a-Lario},
  {Garcia-Reinaldos}, {Gonz{\'a}lez-N{\'u}{\~n}ez}, {Gosset}, {Haigron}, {Halbwachs}, {Hambly}, {Harrison}, {Hatzidimitriou}, {Heiter}, {Hern{\'a}ndez}, {Hestroffer}, {Hodgkin}, {Holl}, {Jan{\ss}en}, {Jevardat de Fombelle}, {Jordan}, {Krone-Martins}, {Lanzafame}, {L{\"o}ffler}, {Lorca}, {Manteiga}, {Marchal}, {Marrese}, {Moitinho}, {Mora}, {Muinonen}, {Osborne}, {Pancino}, {Pauwels}, {Petit}, {Recio-Blanco}, {Richards}, {Riello}, {Rimoldini}, {Robin}, {Roegiers}, {Rybizki}, {Sarro}, {Siopis}, {Smith}, {Sozzetti}, {Ulla}, {Utrilla}, {van Leeuwen}, {van Reeven}, {Abbas}, {Abreu Aramburu}, {Accart}, {Aerts}, {Aguado}, {Ajaj}, {Altavilla}, {{\'A}lvarez}, {{\'A}lvarez Cid-Fuentes}, {Alves}, {Anderson}, {Anglada Varela}, {Antoja}, {Audard}, {Baines}, {Baker}, {Balaguer-N{\'u}{\~n}ez}, {Balbinot}, {Balog}, {Barache}, {Barbato}, {Barros}, {Barstow}, {Bartolom{\'e}}, {Bassilana}, {Bauchet}, {Baudesson-Stella}, {Becciani}, {Bellazzini}, {Bernet}, {Bertone}, {Bianchi}, {Blanco-Cuaresma}, {Boch}, {Bombrun}, {Bossini},
  {Bouquillon}, {Bragaglia}, {Bramante}, {Breedt}, {Bressan}, {Brouillet}, {Bucciarelli}, {Burlacu}, {Busonero}, {Butkevich}, {Buzzi}, {Caffau}, {Cancelliere}, {C{\'a}novas}, {Cantat-Gaudin}, {Carballo}, {Carlucci}, {Carnerero}, {Carrasco}, {Casamiquela}, {Castellani}, {Castro-Ginard}, {Castro Sampol}, {Chaoul}, {Charlot}, {Chemin}, {Chiavassa}, {Cioni}, {Comoretto}, {Cooper}, {Cornez}, {Cowell}, {Crifo}, {Crosta}, {Crowley}, {Dafonte}, {Dapergolas}, {David}, \& {David}}]{Gaia2021A&A...649A...1G}
{Gaia Collaboration}, {Brown}, A.~G.~A., {Vallenari}, A., {et~al.} 2021, \aap, 649, A1, \dodoi{10.1051/0004-6361/202039657}

\bibitem[{{Gaia Collaboration} {et~al.}(2023){Gaia Collaboration}, {Vallenari}, {Brown}, {Prusti}, {de Bruijne}, {Arenou}, {Babusiaux}, {Biermann}, {Creevey}, {Ducourant}, {Evans}, {Eyer}, {Guerra}, {Hutton}, {Jordi}, {Klioner}, {Lammers}, {Lindegren}, {Luri}, {Mignard}, {Panem}, {Pourbaix}, {Randich}, {Sartoretti}, {Soubiran}, {Tanga}, {Walton}, {Bailer-Jones}, {Bastian}, {Drimmel}, {Jansen}, {Katz}, {Lattanzi}, {van Leeuwen}, {Bakker}, {Cacciari}, {Casta{\~n}eda}, {De Angeli}, {Fabricius}, {Fouesneau}, {Fr{\'e}mat}, {Galluccio}, {Guerrier}, {Heiter}, {Masana}, {Messineo}, {Mowlavi}, {Nicolas}, {Nienartowicz}, {Pailler}, {Panuzzo}, {Riclet}, {Roux}, {Seabroke}, {Sordo}, {Th{\'e}venin}, {Gracia-Abril}, {Portell}, {Teyssier}, {Altmann}, {Andrae}, {Audard}, {Bellas-Velidis}, {Benson}, {Berthier}, {Blomme}, {Burgess}, {Busonero}, {Busso}, {C{\'a}novas}, {Carry}, {Cellino}, {Cheek}, {Clementini}, {Damerdji}, {Davidson}, {de Teodoro}, {Nu{\~n}ez Campos}, {Delchambre}, {Dell'Oro}, {Esquej},
  {Fern{\'a}ndez-Hern{\'a}ndez}, {Fraile}, {Garabato}, {Garc{\'\i}a-Lario}, {Gosset}, {Haigron}, {Halbwachs}, {Hambly}, {Harrison}, {Hern{\'a}ndez}, {Hestroffer}, {Hodgkin}, {Holl}, {Jan{\ss}en}, {Jevardat de Fombelle}, {Jordan}, {Krone-Martins}, {Lanzafame}, {L{\"o}ffler}, {Marchal}, {Marrese}, {Moitinho}, {Muinonen}, {Osborne}, {Pancino}, {Pauwels}, {Recio-Blanco}, {Reyl{\'e}}, {Riello}, {Rimoldini}, {Roegiers}, {Rybizki}, {Sarro}, {Siopis}, {Smith}, {Sozzetti}, {Utrilla}, {van Leeuwen}, {Abbas}, {{\'A}brah{\'a}m}, {Abreu Aramburu}, {Aerts}, {Aguado}, {Ajaj}, {Aldea-Montero}, {Altavilla}, {{\'A}lvarez}, {Alves}, {Anders}, {Anderson}, {Anglada Varela}, {Antoja}, {Baines}, {Baker}, {Balaguer-N{\'u}{\~n}ez}, {Balbinot}, {Balog}, {Barache}, {Barbato}, {Barros}, {Barstow}, {Bartolom{\'e}}, {Bassilana}, {Bauchet}, {Becciani}, {Bellazzini}, {Berihuete}, {Bernet}, {Bertone}, {Bianchi}, {Binnenfeld}, {Blanco-Cuaresma}, {Blazere}, {Boch}, {Bombrun}, {Bossini}, {Bouquillon}, {Bragaglia}, {Bramante}, {Breedt},
  {Bressan}, {Brouillet}, {Brugaletta}, {Bucciarelli}, {Burlacu}, {Butkevich}, {Buzzi}, {Caffau}, {Cancelliere}, {Cantat-Gaudin}, {Carballo}, {Carlucci}, {Carnerero}, {Carrasco}, {Casamiquela}, {Castellani}, {Castro-Ginard}, {Chaoul}, {Charlot}, {Chemin}, {Chiaramida}, {Chiavassa}, {Chornay}, {Comoretto}, {Contursi}, {Cooper}, {Cornez}, {Cowell}, {Crifo}, {Cropper}, {Crosta}, {Crowley}, {Dafonte}, {Dapergolas}, {David}, {David}, {de Laverny}, {De Luise}, {De March}, {De Ridder}, {de Souza}, {de Torres}, {del Peloso}, {del Pozo}, {Delbo}, {Delgado}, {Delisle}, {Demouchy}, {Dharmawardena}, {Di Matteo}, {Diakite}, {Diener}, {Distefano}, {Dolding}, {Edvardsson}, {Enke}, {Fabre}, {Fabrizio}, {Faigler}, {Fedorets}, {Fernique}, {Fienga}, {Figueras}, {Fournier}, {Fouron}, {Fragkoudi}, {Gai}, {Garcia-Gutierrez}, {Garcia-Reinaldos}, {Garc{\'\i}a-Torres}, {Garofalo}, {Gavel}, {Gavras}, {Gerlach}, {Geyer}, {Giacobbe}, {Gilmore}, {Girona}, {Giuffrida}, {Gomel}, {Gomez}, {Gonz{\'a}lez-N{\'u}{\~n}ez},
  {Gonz{\'a}lez-Santamar{\'\i}a}, {Gonz{\'a}lez-Vidal}, {Granvik}, {Guillout}, {Guiraud}, {Guti{\'e}rrez-S{\'a}nchez}, {Guy}, {Hatzidimitriou}, {Hauser}, {Haywood}, {Helmer}, {Helmi}, {Sarmiento}, {Hidalgo}, {Hilger}, {H{\l}adczuk}, {Hobbs}, {Holland}, {Huckle}, {Jardine}, {Jasniewicz}, {Jean-Antoine Piccolo}, {Jim{\'e}nez-Arranz}, {Jorissen}, {Juaristi Campillo}, {Julbe}, {Karbevska}, {Kervella}, {Khanna}, {Kontizas}, {Kordopatis}, {Korn}, {K{\'o}sp{\'a}l}, {Kostrzewa-Rutkowska}, {Kruszy{\'n}ska}, {Kun}, {Laizeau}, {Lambert}, {Lanza}, {Lasne}, {Le Campion}, {Lebreton}, {Lebzelter}, {Leccia}, {Leclerc}, {Lecoeur-Taibi}, {Liao}, {Licata}, {Lindstr{\o}m}, {Lister}, {Livanou}, {Lobel}, {Lorca}, {Loup}, {Madrero Pardo}, {Magdaleno Romeo}, {Managau}, {Mann}, {Manteiga}, {Marchant}, {Marconi}, {Marcos}, {Marcos Santos}, {Mar{\'\i}n Pina}, {Marinoni}, {Marocco}, {Marshall}, {Martin Polo}, {Mart{\'\i}n-Fleitas}, {Marton}, {Mary}, {Masip}, {Massari}, {Mastrobuono-Battisti}, {Mazeh}, {McMillan}, {Messina}, {Michalik},
  {Millar}, {Mints}, {Molina}, {Molinaro}, {Moln{\'a}r}, {Monari}, {Mongui{\'o}}, {Montegriffo}, {Montero}, {Mor}, {Mora}, {Morbidelli}, {Morel}, {Morris}, {Muraveva}, {Murphy}, {Musella}, {Nagy}, {Noval}, {Oca{\~n}a}, {Ogden}, {Ordenovic}, {Osinde}, {Pagani}, {Pagano}, {Palaversa}, {Palicio}, {Pallas-Quintela}, {Panahi}, {Payne-Wardenaar}, {Pe{\~n}alosa Esteller}, {Penttil{\"a}}, {Pichon}, {Piersimoni}, {Pineau}, {Plachy}, {Plum}, {Poggio}, {Pr{\v{s}}a}, {Pulone}, {Racero}, {Ragaini}, {Rainer}, {Raiteri}, {Rambaux}, {Ramos}, {Ramos-Lerate}, {Re Fiorentin}, {Regibo}, {Richards}, {Rios Diaz}, {Ripepi}, {Riva}, {Rix}, {Rixon}, {Robichon}, {Robin}, {Robin}, {Roelens}, {Rogues}, {Rohrbasser}, {Romero-G{\'o}mez}, {Rowell}, {Royer}, {Ruz Mieres}, {Rybicki}, {Sadowski}, {S{\'a}ez N{\'u}{\~n}ez}, {Sagrist{\`a} Sell{\'e}s}, {Sahlmann}, {Salguero}, {Samaras}, {Sanchez Gimenez}, {Sanna}, {Santove{\~n}a}, {Sarasso}, {Schultheis}, {Sciacca}, {Segol}, {Segovia}, {S{\'e}gransan}, {Semeux}, {Shahaf}, {Siddiqui}, {Siebert},
  {Siltala}, {Silvelo}, {Slezak}, {Slezak}, {Smart}, {Snaith}, {Solano}, {Solitro}, {Souami}, {Souchay}, {Spagna}, {Spina}, {Spoto}, {Steele}, {Steidelm{\"u}ller}, {Stephenson}, {S{\"u}veges}, {Surdej}, {Szabados}, {Szegedi-Elek}, {Taris}, {Taylor}, {Teixeira}, {Tolomei}, {Tonello}, {Torra}, {Torra}, {Torralba Elipe}, {Trabucchi}, {Tsounis}, {Turon}, {Ulla}, {Unger}, {Vaillant}, {van Dillen}, {van Reeven}, {Vanel}, {Vecchiato}, {Viala}, {Vicente}, {Voutsinas}, {Weiler}, {Wevers}, {Wyrzykowski}, {Yoldas}, {Yvard}, {Zhao}, {Zorec}, {Zucker}, \& {Zwitter}}]{Gaia2023}
{Gaia Collaboration}, {Vallenari}, A., {Brown}, A.~G.~A., {et~al.} 2023, \aap, 674, A1, \dodoi{10.1051/0004-6361/202243940}

\bibitem[{{Gan}(2023)}]{Gan2023gaia}
{Gan}, T. 2023, Research Notes of the American Astronomical Society, 7, 226, \dodoi{10.3847/2515-5172/ad0643}

\bibitem[{{Gan} {et~al.}(2023){Gan}, {Cadieux}, {Jahandar}, {Vazan}, {Wang}, {Mao}, {Alvarado-Montes}, {Lin}, {Artigau}, {Cook}, {Doyon}, {Mann}, {Stassun}, {Burgasser}, {Rackham}, {Howell}, {Collins}, {Barkaoui}, {Shporer}, {de Leon}, {Arnold}, {Ricker}, {Vanderspek}, {Latham}, {Seager}, {Winn}, {Jenkins}, {Burdanov}, {Charbonneau}, {Dransfield}, {Fukui}, {Furlan}, {Gillon}, {Hooton}, {Lewis}, {Littlefield}, {Mireles}, {Narita}, {Ormel}, {Quinn}, {Sefako}, {Timmermans}, {Vezie}, \& {de Wit}}]{Gan2023TOI4201}
{Gan}, T., {Cadieux}, C., {Jahandar}, F., {et~al.} 2023, \aj, 166, 165, \dodoi{10.3847/1538-3881/acf56d}

\bibitem[{{Giacalone} {et~al.}(2024){Giacalone}, {Dai}, {Zanazzi}, {Howard}, {Dressing}, {Winn}, {Rubenzahl}, {Carmichael}, {Vowell}, {Kesseli}, {Halverson}, {Isaacson}, {Brodheim}, {Deich}, {Fulton}, {Gibson}, {Hill}, {Holden}, {Householder}, {Kaye}, {Laher}, {Lanclos}, {Payne}, {Petigura}, {Roy}, {Schwab}, {Shaum}, {Sirk}, {Smith}, {Stef{\'a}nsson}, {Walawender}, {Wang}, {Weiss}, \& {Yeh}}]{Giacalone2024}
{Giacalone}, S., {Dai}, F., {Zanazzi}, J.~J., {et~al.} 2024, \aj, 168, 189, \dodoi{10.3847/1538-3881/ad785a}

\bibitem[{{Goldreich} \& {Sari}(2003)}]{Goldreich2003}
{Goldreich}, P., \& {Sari}, R. 2003, \apj, 585, 1024, \dodoi{10.1086/346202}

\bibitem[{{Goldreich} \& {Soter}(1966)}]{Goldreich1966}
{Goldreich}, P., \& {Soter}, S. 1966, \icarus, 5, 375, \dodoi{10.1016/0019-1035(66)90051-0}

\bibitem[{{Grether} \& {Lineweaver}(2006)}]{Grether2006}
{Grether}, D., \& {Lineweaver}, C.~H. 2006, \apj, 640, 1051, \dodoi{10.1086/500161}

\bibitem[{{Halbwachs} {et~al.}(2023){Halbwachs}, {Pourbaix}, {Arenou}, {Galluccio}, {Guillout}, {Bauchet}, {Marchal}, {Sadowski}, \& {Teyssier}}]{Halbwachs2023}
{Halbwachs}, J.-L., {Pourbaix}, D., {Arenou}, F., {et~al.} 2023, \aap, 674, A9, \dodoi{10.1051/0004-6361/202243969}

\bibitem[{{Heggie} \& {Rasio}(1996)}]{Heggie1996}
{Heggie}, D.~C., \& {Rasio}, F.~A. 1996, \mnras, 282, 1064, \dodoi{10.1093/mnras/282.3.1064}

\bibitem[{{Henderson} {et~al.}(2024){Henderson}, {Casewell}, {Jord{\'a}n}, {Brahm}, {Henning}, {Gill}, {Mayorga}, {Ziegler}, {Stassun}, {Goad}, {Acton}, {Alves}, {Anderson}, {Apergis}, {Armstrong}, {Bayliss}, {Burleigh}, {Dragomir}, {Gillen}, {G{\"u}nther}, {Hedges}, {Hesse}, {Hobson}, {Jenkins}, {Jenkins}, {Kendall}, {Lendl}, {Lund}, {McCormac}, {Moyano}, {Osborn}, {Pinto}, {Ramsay}, {Rapetti}, {Saha}, {Seager}, {Trifonov}, {Udry}, {Vines}, {West}, {Wheatley}, {Winn}, \& {Zivave}}]{Henderson2024}
{Henderson}, B.~A., {Casewell}, S.~L., {Jord{\'a}n}, A., {et~al.} 2024, \mnras, 533, 2823, \dodoi{10.1093/mnras/stae1940}

\bibitem[{{Hogg} {et~al.}(2010){Hogg}, {Myers}, \& {Bovy}}]{Hogg2010}
{Hogg}, D.~W., {Myers}, A.~D., \& {Bovy}, J. 2010, \apj, 725, 2166, \dodoi{10.1088/0004-637X/725/2/2166}

\bibitem[{{Holl} {et~al.}(2023){Holl}, {Sozzetti}, {Sahlmann}, {Giacobbe}, {S{\'e}gransan}, {Unger}, {Delisle}, {Barbato}, {Lattanzi}, {Morbidelli}, \& {Sosnowska}}]{Holl2023}
{Holl}, B., {Sozzetti}, A., {Sahlmann}, J., {et~al.} 2023, \aap, 674, A10, \dodoi{10.1051/0004-6361/202244161}

\bibitem[{{Howell} {et~al.}(2016){Howell}, {Everett}, {Horch}, {Winters}, {Hirsch}, {Nusdeo}, \& {Scott}}]{Howell2016}
{Howell}, S.~B., {Everett}, M.~E., {Horch}, E.~P., {et~al.} 2016, \apjl, 829, L2, \dodoi{10.3847/2041-8205/829/1/L2}

\bibitem[{{Howell} {et~al.}(2011){Howell}, {Everett}, {Sherry}, {Horch}, \& {Ciardi}}]{Howell2011}
{Howell}, S.~B., {Everett}, M.~E., {Sherry}, W., {Horch}, E., \& {Ciardi}, D.~R. 2011, \aj, 142, 19, \dodoi{10.1088/0004-6256/142/1/19}

\bibitem[{{Huang} {et~al.}(2020{\natexlab{a}}){Huang}, {Vanderburg}, {P{\'a}l}, {Sha}, {Yu}, {Fong}, {Fausnaugh}, {Shporer}, {Guerrero}, {Vanderspek}, \& {Ricker}}]{QLP2020a}
{Huang}, C.~X., {Vanderburg}, A., {P{\'a}l}, A., {et~al.} 2020{\natexlab{a}}, Research Notes of the American Astronomical Society, 4, 204, \dodoi{10.3847/2515-5172/abca2e}

\bibitem[{{Huang} {et~al.}(2020{\natexlab{b}}){Huang}, {Vanderburg}, {P{\'a}l}, {Sha}, {Yu}, {Fong}, {Fausnaugh}, {Shporer}, {Guerrero}, {Vanderspek}, \& {Ricker}}]{QLP2020b}
---. 2020{\natexlab{b}}, Research Notes of the American Astronomical Society, 4, 206, \dodoi{10.3847/2515-5172/abca2d}

\bibitem[{Husser {et~al.}(2013)Husser, {Wende-von Berg}, Dreizler, Homeier, Reiners, Barman, \& Hauschildt}]{Husser2013}
Husser, T.-O., {Wende-von Berg}, S., Dreizler, S., {et~al.} 2013, A{\&}A, 553, A6, \dodoi{10.1051/0004-6361/201219058}

\bibitem[{{Ida} {et~al.}(2013){Ida}, {Lin}, \& {Nagasawa}}]{Ida2013}
{Ida}, S., {Lin}, D.~N.~C., \& {Nagasawa}, M. 2013, \apj, 775, 42, \dodoi{10.1088/0004-637X/775/1/42}

\bibitem[{{Ida} {et~al.}(2020){Ida}, {Muto}, {Matsumura}, \& {Brasser}}]{Ida2020}
{Ida}, S., {Muto}, T., {Matsumura}, S., \& {Brasser}, R. 2020, \mnras, 494, 5666, \dodoi{10.1093/mnras/staa1073}

\bibitem[{{Jackson} {et~al.}(2008){Jackson}, {Greenberg}, \& {Barnes}}]{Jackson2008}
{Jackson}, B., {Greenberg}, R., \& {Barnes}, R. 2008, \apj, 678, 1396, \dodoi{10.1086/529187}

\bibitem[{{Jenkins} {et~al.}(2016){Jenkins}, {Twicken}, {McCauliff}, {Campbell}, {Sanderfer}, {Lung}, {Mansouri-Samani}, {Girouard}, {Tenenbaum}, {Klaus}, {Smith}, {Caldwell}, {Chacon}, {Henze}, {Heiges}, {Latham}, {Morgan}, {Swade}, {Rinehart}, \& {Vanderspek}}]{Jenkins2016}
{Jenkins}, J.~M., {Twicken}, J.~D., {McCauliff}, S., {et~al.} 2016, in \procspie, Vol. 9913, Software and Cyberinfrastructure for Astronomy IV, 99133E, \dodoi{10.1117/12.2233418}

\bibitem[{{Kipping}(2013)}]{Kipping2013}
{Kipping}, D.~M. 2013, \mnras, 435, 2152, \dodoi{10.1093/mnras/stt1435}

\bibitem[{{Kratter} \& {Lodato}(2016)}]{Kratter2016}
{Kratter}, K., \& {Lodato}, G. 2016, \araa, 54, 271, \dodoi{10.1146/annurev-astro-081915-023307}

\bibitem[{{Kreidberg}(2015)}]{Kreidberg2015}
{Kreidberg}, L. 2015, \pasp, 127, 1161, \dodoi{10.1086/683602}

\bibitem[{{Kunimoto} {et~al.}(2022){Kunimoto}, {Daylan}, {Guerrero}, {Fong}, {Bryson}, {Ricker}, {Fausnaugh}, {Huang}, {Sha}, {Shporer}, {Vanderburg}, {Vanderspek}, \& {Yu}}]{Kunimoto2022}
{Kunimoto}, M., {Daylan}, T., {Guerrero}, N., {et~al.} 2022, \apjs, 259, 33, \dodoi{10.3847/1538-4365/ac5688}

\bibitem[{{Lai} \& {Mu{\~n}oz}(2023)}]{Lai2023}
{Lai}, D., \& {Mu{\~n}oz}, D.~J. 2023, \araa, 61, 517, \dodoi{10.1146/annurev-astro-052622-022933}

\bibitem[{{Larsen} {et~al.}(2025){Larsen}, {Swaby}, {Kobulnicky}, {Ca{\~n}as}, {Kanodia}, {Libby-Roberts}, {Monson}, {Gupta}, {Cochran}, {Mahadevan}, {Bender}, {Diddams}, {Halverson}, {Lin}, {Moe}, {Ninan}, {Robertson}, {Roy}, {Schwab}, \& {Stefansson}}]{Larsen2025}
{Larsen}, A., {Swaby}, T.~N., {Kobulnicky}, H.~A., {et~al.} 2025, \aj, 169, 246, \dodoi{10.3847/1538-3881/adbb54}

\bibitem[{{Laughlin} {et~al.}(1997){Laughlin}, {Bodenheimer}, \& {Adams}}]{Laughlin1997}
{Laughlin}, G., {Bodenheimer}, P., \& {Adams}, F.~C. 1997, \apj, 482, 420, \dodoi{10.1086/304125}

\bibitem[{{Lightkurve Collaboration} {et~al.}(2018){Lightkurve Collaboration}, {Cardoso}, {Hedges}, {Gully-Santiago}, {Saunders}, {Cody}, {Barclay}, {Hall}, {Sagear}, {Turtelboom}, {Zhang}, {Tzanidakis}, {Mighell}, {Coughlin}, {Bell}, {Berta-Thompson}, {Williams}, {Dotson}, \& {Barentsen}}]{lightkurvecollaboration}
{Lightkurve Collaboration}, {Cardoso}, J. V. d. M.~a., {Hedges}, C., {et~al.} 2018, {Lightkurve: Kepler and TESS time series analysis in Python}.
\newblock \doeprint{1812.013}

\bibitem[{{Lin} {et~al.}(1996){Lin}, {Bodenheimer}, \& {Richardson}}]{Lin1996}
{Lin}, D.~N.~C., {Bodenheimer}, P., \& {Richardson}, D.~C. 1996, \nat, 380, 606, \dodoi{10.1038/380606a0}

\bibitem[{{Lin} {et~al.}(2023){Lin}, {Gan}, {Wang}, {Shporer}, {Rabus}, {Zhou}, {Psaridi}, {Bouchy}, {Bieryla}, {Latham}, {Mao}, {Stassun}, {Hellier}, {Howell}, {Ziegler}, {Caldwell}, {Clark}, {Collins}, {Curtis}, {Faherty}, {Gnilka}, {Grunblatt}, {Jenkins}, {Johnson}, {Law}, {Lendl}, {Littlefield}, {Lund}, {Lund}, {Mann}, {McDermott}, {Mishra}, {Mounzer}, {Paegert}, {Pritchard}, {Ricker}, {Seager}, {Srdoc}, {Sun}, {Tang}, {Udry}, {Vanderspek}, {Watanabe}, {Winn}, \& {Yu}}]{Lin2023}
{Lin}, Z., {Gan}, T., {Wang}, S.~X., {et~al.} 2023, \mnras, 523, 6162, \dodoi{10.1093/mnras/stad1745}

\bibitem[{{Ma} \& {Ge}(2014)}]{Ma2014}
{Ma}, B., \& {Ge}, J. 2014, \mnras, 439, 2781, \dodoi{10.1093/mnras/stu134}

\bibitem[{{Mann} {et~al.}(2014){Mann}, {Deacon}, {Gaidos}, {Ansdell}, {Brewer}, {Liu}, {Magnier}, \& {Aller}}]{2014AJ....147..160M}
{Mann}, A.~W., {Deacon}, N.~R., {Gaidos}, E., {et~al.} 2014, \aj, 147, 160, \dodoi{10.1088/0004-6256/147/6/160}

\bibitem[{{Mann} {et~al.}(2015){Mann}, {Feiden}, {Gaidos}, {Boyajian}, \& {von Braun}}]{Mann2015}
{Mann}, A.~W., {Feiden}, G.~A., {Gaidos}, E., {Boyajian}, T., \& {von Braun}, K. 2015, \apj, 804, 64, \dodoi{10.1088/0004-637X/804/1/64}

\bibitem[{{Mann} {et~al.}(2019){Mann}, {Dupuy}, {Kraus}, {Gaidos}, {Ansdell}, {Ireland}, {Rizzuto}, {Hung}, {Dittmann}, {Factor}, {Feiden}, {Martinez}, {Ru{\'\i}z-Rodr{\'\i}guez}, \& {Thao}}]{Mann2019}
{Mann}, A.~W., {Dupuy}, T., {Kraus}, A.~L., {et~al.} 2019, \apj, 871, 63, \dodoi{10.3847/1538-4357/aaf3bc}

\bibitem[{{Marcy} \& {Butler}(2000)}]{Marcy2000}
{Marcy}, G.~W., \& {Butler}, R.~P. 2000, \pasp, 112, 137, \dodoi{10.1086/316516}

\bibitem[{{Masci} {et~al.}(2019){Masci}, {Laher}, {Rusholme}, {Shupe}, {Groom}, {Surace}, {Jackson}, {Monkewitz}, {Beck}, {Flynn}, {Terek}, {Landry}, {Hacopians}, {Desai}, {Howell}, {Brooke}, {Imel}, {Wachter}, {Ye}, {Lin}, {Cenko}, {Cunningham}, {Rebbapragada}, {Bue}, {Miller}, {Mahabal}, {Bellm}, {Patterson}, {Juri{\'c}}, {Golkhou}, {Ofek}, {Walters}, {Graham}, {Kasliwal}, {Dekany}, {Kupfer}, {Burdge}, {Cannella}, {Barlow}, {Van Sistine}, {Giomi}, {Fremling}, {Blagorodnova}, {Levitan}, {Riddle}, {Smith}, {Helou}, {Prince}, \& {Kulkarni}}]{Masci2019}
{Masci}, F.~J., {Laher}, R.~R., {Rusholme}, B., {et~al.} 2019, \pasp, 131, 018003, \dodoi{10.1088/1538-3873/aae8ac}

\bibitem[{{Nagpal} {et~al.}(2023){Nagpal}, {Blunt}, {Bowler}, {Dupuy}, {Nielsen}, \& {Wang}}]{Nagpal2023}
{Nagpal}, V., {Blunt}, S., {Bowler}, B.~P., {et~al.} 2023, \aj, 165, 32, \dodoi{10.3847/1538-3881/ac9fd2}

\bibitem[{{Naoz} \& {Fabrycky}(2014)}]{Naoz2014}
{Naoz}, S., \& {Fabrycky}, D.~C. 2014, \apj, 793, 137, \dodoi{10.1088/0004-637X/793/2/137}

\bibitem[{{Naoz} {et~al.}(2011){Naoz}, {Farr}, {Lithwick}, {Rasio}, \& {Teyssandier}}]{Naoz2011}
{Naoz}, S., {Farr}, W.~M., {Lithwick}, Y., {Rasio}, F.~A., \& {Teyssandier}, J. 2011, \nat, 473, 187, \dodoi{10.1038/nature10076}

\bibitem[{{Newton} {et~al.}(2014){Newton}, {Charbonneau}, {Irwin}, {Berta-Thompson}, {Rojas-Ayala}, {Covey}, \& {Lloyd}}]{2014AJ....147...20N}
{Newton}, E.~R., {Charbonneau}, D., {Irwin}, J., {et~al.} 2014, \aj, 147, 20, \dodoi{10.1088/0004-6256/147/1/20}

\bibitem[{{Pecaut} \& {Mamajek}(2013)}]{Pecaut2013}
{Pecaut}, M.~J., \& {Mamajek}, E.~E. 2013, \apjs, 208, 9, \dodoi{10.1088/0067-0049/208/1/9}

\bibitem[{{Petrovich} \& {Tremaine}(2016)}]{Petrovich2016}
{Petrovich}, C., \& {Tremaine}, S. 2016, \apj, 829, 132, \dodoi{10.3847/0004-637X/829/2/132}

\bibitem[{{Pollack} {et~al.}(1996){Pollack}, {Hubickyj}, {Bodenheimer}, {Lissauer}, {Podolak}, \& {Greenzweig}}]{Pollack1996}
{Pollack}, J.~B., {Hubickyj}, O., {Bodenheimer}, P., {et~al.} 1996, \icarus, 124, 62, \dodoi{10.1006/icar.1996.0190}

\bibitem[{{Psaridi} {et~al.}(2022){Psaridi}, {Bouchy}, {Lendl}, {Grieves}, {Stassun}, {Carmichael}, {Gill}, {Pe{\~n}a Rojas}, {Gan}, {Shporer}, {Bieryla}, {Brahm}, {Christiansen}, {Crossfield}, {Galland}, {Hooton}, {Jenkins}, {Jenkins}, {Latham}, {Lund}, {Rodriguez}, {Ting}, {Udry}, {Ulmer-Moll}, {Wittenmyer}, {Zhang}, {Zhou}, {Addison}, {Cointepas}, {Collins}, {Collins}, {Deline}, {Dressing}, {Evans}, {Giacalone}, {Heitzmann}, {Mireles}, {Mounzer}, {Otegi}, {Radford}, {Rudat}, {Schlieder}, {Schwarz}, {Srdoc}, {Stockdale}, {Suarez}, {Wright}, \& {Zhao}}]{Psaridi2022}
{Psaridi}, A., {Bouchy}, F., {Lendl}, M., {et~al.} 2022, \aap, 664, A94, \dodoi{10.1051/0004-6361/202243454}

\bibitem[{{Rasio} \& {Ford}(1996)}]{Rasio1996}
{Rasio}, F.~A., \& {Ford}, E.~B. 1996, Science, 274, 954, \dodoi{10.1126/science.274.5289.954}

\bibitem[{{Ricker} {et~al.}(2015){Ricker}, {Winn}, {Vanderspek}, {Latham}, {Bakos}, {Bean}, {Berta-Thompson}, {Brown}, {Buchhave}, {Butler}, {Butler}, {Chaplin}, {Charbonneau}, {Christensen-Dalsgaard}, {Clampin}, {Deming}, {Doty}, {De Lee}, {Dressing}, {Dunham}, {Endl}, {Fressin}, {Ge}, {Henning}, {Holman}, {Howard}, {Ida}, {Jenkins}, {Jernigan}, {Johnson}, {Kaltenegger}, {Kawai}, {Kjeldsen}, {Laughlin}, {Levine}, {Lin}, {Lissauer}, {MacQueen}, {Marcy}, {McCullough}, {Morton}, {Narita}, {Paegert}, {Palle}, {Pepe}, {Pepper}, {Quirrenbach}, {Rinehart}, {Sasselov}, {Sato}, {Seager}, {Sozzetti}, {Stassun}, {Sullivan}, {Szentgyorgyi}, {Torres}, {Udry}, \& {Villasenor}}]{Ricker2015}
{Ricker}, G.~R., {Winn}, J.~N., {Vanderspek}, R., {et~al.} 2015, Journal of Astronomical Telescopes, Instruments, and Systems, 1, 014003, \dodoi{10.1117/1.JATIS.1.1.014003}

\bibitem[{{Rojas-Ayala} {et~al.}(2012){Rojas-Ayala}, {Covey}, {Muirhead}, \& {Lloyd}}]{2012ApJ...748...93R}
{Rojas-Ayala}, B., {Covey}, K.~R., {Muirhead}, P.~S., \& {Lloyd}, J.~P. 2012, \apj, 748, 93, \dodoi{10.1088/0004-637X/748/2/93}

\bibitem[{{Scott} {et~al.}(2021){Scott}, {Howell}, {Gnilka}, {Stephens}, {Salinas}, {Matson}, {Furlan}, {Horch}, {Everett}, {Ciardi}, {Mills}, \& {Quigley}}]{Scott2021}
{Scott}, N.~J., {Howell}, S.~B., {Gnilka}, C.~L., {et~al.} 2021, Frontiers in Astronomy and Space Sciences, 8, 138, \dodoi{10.3389/fspas.2021.716560}

\bibitem[{{Siverd} {et~al.}(2012){Siverd}, {Beatty}, {Pepper}, {Eastman}, {Collins}, {Bieryla}, {Latham}, {Buchhave}, {Jensen}, {Crepp}, {Street}, {Stassun}, {Gaudi}, {Berlind}, {Calkins}, {DePoy}, {Esquerdo}, {Fulton}, {F{\H{u}}r{\'e}sz}, {Geary}, {Gould}, {Hebb}, {Kielkopf}, {Marshall}, {Pogge}, {Stanek}, {Stefanik}, {Szentgyorgyi}, {Trueblood}, {Trueblood}, {Stutz}, \& {van Saders}}]{Siverd2012}
{Siverd}, R.~J., {Beatty}, T.~G., {Pepper}, J., {et~al.} 2012, \apj, 761, 123, \dodoi{10.1088/0004-637X/761/2/123}

\bibitem[{{Siwek} {et~al.}(2023){Siwek}, {Weinberger}, \& {Hernquist}}]{Siwek2023}
{Siwek}, M., {Weinberger}, R., \& {Hernquist}, L. 2023, \mnras, 522, 2707, \dodoi{10.1093/mnras/stad1131}

\bibitem[{{Smith} {et~al.}(2012){Smith}, {Stumpe}, {Van Cleve}, {Jenkins}, {Barclay}, {Fanelli}, {Girouard}, {Kolodziejczak}, {McCauliff}, {Morris}, \& {Twicken}}]{Smith2012}
{Smith}, J.~C., {Stumpe}, M.~C., {Van Cleve}, J.~E., {et~al.} 2012, \pasp, 124, 1000, \dodoi{10.1086/667697}

\bibitem[{{Speagle}(2020)}]{Speagle2019}
{Speagle}, J.~S. 2020, \mnras, 493, 3132, \dodoi{10.1093/mnras/staa278}

\bibitem[{{Spiegel} {et~al.}(2011){Spiegel}, {Burrows}, \& {Milsom}}]{Spiegel2011}
{Spiegel}, D.~S., {Burrows}, A., \& {Milsom}, J.~A. 2011, \apj, 727, 57, \dodoi{10.1088/0004-637X/727/1/57}

\bibitem[{{Stassun} {et~al.}(2017){Stassun}, {Collins}, \& {Gaudi}}]{Stassun:2017}
{Stassun}, K.~G., {Collins}, K.~A., \& {Gaudi}, B.~S. 2017, \aj, 153, 136, \dodoi{10.3847/1538-3881/aa5df3}

\bibitem[{{Stassun} \& {Torres}(2016)}]{Stassun:2016}
{Stassun}, K.~G., \& {Torres}, G. 2016, \aj, 152, 180, \dodoi{10.3847/0004-6256/152/6/180}

\bibitem[{{Stassun} \& {Torres}(2018)}]{Stassun:2018}
---. 2018, \apj, 862, 61, \dodoi{10.3847/1538-4357/aacafc}

\bibitem[{{Stassun} {et~al.}(2019){Stassun}, {Oelkers}, {Paegert}, {Torres}, {Pepper}, {De Lee}, {Collins}, {Latham}, {Muirhead}, {Chittidi}, {Rojas-Ayala}, {Fleming}, {Rose}, {Tenenbaum}, {Ting}, {Kane}, {Barclay}, {Bean}, {Brassuer}, {Charbonneau}, {Ge}, {Lissauer}, {Mann}, {McLean}, {Mullally}, {Narita}, {Plavchan}, {Ricker}, {Sasselov}, {Seager}, {Sharma}, {Shiao}, {Sozzetti}, {Stello}, {Vanderspek}, {Wallace}, \& {Winn}}]{Stassun2019tic}
{Stassun}, K.~G., {Oelkers}, R.~J., {Paegert}, M., {et~al.} 2019, \aj, 158, 138, \dodoi{10.3847/1538-3881/ab3467}

\bibitem[{{Stef{\'a}nsson} {et~al.}(2025){Stef{\'a}nsson}, {Mahadevan}, {Winn}, {Marcussen}, {Kanodia}, {Albrecht}, {Fitzmaurice}, {Mikulskyt{\.{e}}}, {Ca{\~n}as}, {Espinoza-Retamal}, {Zwart}, {Krolikowski}, {Hotnisky}, {Robertson}, {Alvarado-Montes}, {Bender}, {Blake}, {Callingham}, {Cochran}, {Delamer}, {Diddams}, {Dong}, {Fernandes}, {Giovinazzi}, {Halverson}, {Libby-Roberts}, {Logsdon}, {McElwain}, {Ninan}, {Rajagopal}, {Reji}, {Roy}, {Schwab}, \& {Wright}}]{Stefansson2025}
{Stef{\'a}nsson}, G., {Mahadevan}, S., {Winn}, J.~N., {et~al.} 2025, \aj, 169, 107, \dodoi{10.3847/1538-3881/ada9e1}

\bibitem[{{Stevenson} {et~al.}(2025){Stevenson}, {Haswell}, {Faria}, {Barnes}, {Barstow}, {Dickinson}, \& {Standing}}]{Stevenson2025}
{Stevenson}, A.~T., {Haswell}, C.~A., {Faria}, J.~P., {et~al.} 2025, \mnras, 539, 727, \dodoi{10.1093/mnras/staf502}

\bibitem[{{Strakhov} {et~al.}(2023){Strakhov}, {Safonov}, \& {Cheryasov}}]{Strakhov2023}
{Strakhov}, I.~A., {Safonov}, B.~S., \& {Cheryasov}, D.~V. 2023, Astrophysical Bulletin, 78, 234, \dodoi{10.1134/S1990341323020104}

\bibitem[{{Stumpe} {et~al.}(2014){Stumpe}, {Smith}, {Catanzarite}, {Van Cleve}, {Jenkins}, {Twicken}, \& {Girouard}}]{Stumpe2014}
{Stumpe}, M.~C., {Smith}, J.~C., {Catanzarite}, J.~H., {et~al.} 2014, \pasp, 126, 100, \dodoi{10.1086/674989}

\bibitem[{{Stumpe} {et~al.}(2012){Stumpe}, {Smith}, {Van Cleve}, {Twicken}, {Barclay}, {Fanelli}, {Girouard}, {Jenkins}, {Kolodziejczak}, {McCauliff}, \& {Morris}}]{Stumpe2012}
{Stumpe}, M.~C., {Smith}, J.~C., {Van Cleve}, J.~E., {et~al.} 2012, \pasp, 124, 985, \dodoi{10.1086/667698}

\bibitem[{{Terrien} {et~al.}(2012){Terrien}, {Mahadevan}, {Bender}, {Deshpande}, {Ramsey}, \& {Bochanski}}]{2012ApJ...747L..38T}
{Terrien}, R.~C., {Mahadevan}, S., {Bender}, C.~F., {et~al.} 2012, \apjl, 747, L38, \dodoi{10.1088/2041-8205/747/2/L38}

\bibitem[{{Teyssandier} \& {Lai}(2020)}]{Teyssandier2020}
{Teyssandier}, J., \& {Lai}, D. 2020, \mnras, 495, 3920, \dodoi{10.1093/mnras/staa1363}

\bibitem[{{Tokovinin} \& {Moe}(2020)}]{Tokovinin2020}
{Tokovinin}, A., \& {Moe}, M. 2020, \mnras, 491, 5158, \dodoi{10.1093/mnras/stz3299}

\bibitem[{{Triaud} {et~al.}(2009){Triaud}, {Queloz}, {Bouchy}, {Moutou}, {Collier Cameron}, {Claret}, {Barge}, {Benz}, {Deleuil}, {Guillot}, {H{\'e}brard}, {Lecavelier Des {\'E}tangs}, {Lovis}, {Mayor}, {Pepe}, \& {Udry}}]{Triaud2009}
{Triaud}, A.~H.~M.~J., {Queloz}, D., {Bouchy}, F., {et~al.} 2009, \aap, 506, 377, \dodoi{10.1051/0004-6361/200911897}

\bibitem[{{Triaud} {et~al.}(2013){Triaud}, {Hebb}, {Anderson}, {Cargile}, {Collier Cameron}, {Doyle}, {Faedi}, {Gillon}, {Gomez Maqueo Chew}, {Hellier}, {Jehin}, {Maxted}, {Naef}, {Pepe}, {Pollacco}, {Queloz}, {S{\'e}gransan}, {Smalley}, {Stassun}, {Udry}, \& {West}}]{Triaud2013}
{Triaud}, A.~H.~M.~J., {Hebb}, L., {Anderson}, D.~R., {et~al.} 2013, \aap, 549, A18, \dodoi{10.1051/0004-6361/201219643}

\bibitem[{{Unger} {et~al.}(2023){Unger}, {S{\'e}gransan}, {Barbato}, {Delisle}, {Sahlmann}, {Holl}, \& {Udry}}]{Unger2023}
{Unger}, N., {S{\'e}gransan}, D., {Barbato}, D., {et~al.} 2023, \aap, 680, A16, \dodoi{10.1051/0004-6361/202347578}

\bibitem[{{Vacca} {et~al.}(2003){Vacca}, {Cushing}, \& {Rayner}}]{2003PASP..115..389V}
{Vacca}, W.~D., {Cushing}, M.~C., \& {Rayner}, J.~T. 2003, \pasp, 115, 389, \dodoi{10.1086/346193}

\bibitem[{{Valli} {et~al.}(2024){Valli}, {Tiede}, {Vigna-G{\'o}mez}, {Cuadra}, {Siwek}, {Ma}, {D'Orazio}, {Zrake}, \& {de Mink}}]{Valli2024}
{Valli}, R., {Tiede}, C., {Vigna-G{\'o}mez}, A., {et~al.} 2024, \aap, 688, A128, \dodoi{10.1051/0004-6361/202449421}

\bibitem[{{Van Eylen} {et~al.}(2019){Van Eylen}, {Albrecht}, {Huang}, {MacDonald}, {Dawson}, {Cai}, {Foreman-Mackey}, {Lundkvist}, {Silva Aguirre}, {Snellen}, \& {Winn}}]{VanEylen2019}
{Van Eylen}, V., {Albrecht}, S., {Huang}, X., {et~al.} 2019, \aj, 157, 61, \dodoi{10.3847/1538-3881/aaf22f}

\bibitem[{{Virtanen} {et~al.}(2020){Virtanen}, {Gommers}, {Oliphant}, {Haberland}, {Reddy}, {Cournapeau}, {Burovski}, {Peterson}, {Weckesser}, {Bright}, {van der Walt}, {Brett}, {Wilson}, {Millman}, {Mayorov}, {Nelson}, {Jones}, {Kern}, {Larson}, {Carey}, {Polat}, {Feng}, {Moore}, {VanderPlas}, {Laxalde}, {Perktold}, {Cimrman}, {Henriksen}, {Quintero}, {Harris}, {Archibald}, {Ribeiro}, {Pedregosa}, {van Mulbregt}, \& {SciPy 1. 0 Contributors}}]{Virtanen2020}
{Virtanen}, P., {Gommers}, R., {Oliphant}, T.~E., {et~al.} 2020, Nature Methods, 17, 261, \dodoi{10.1038/s41592-019-0686-2}

\bibitem[{{Vowell} {et~al.}(2025){Vowell}, {Rodriguez}, {Latham}, {Quinn}, {Schulte}, {Eastman}, {Bieryla}, {Barkaoui}, {Ciardi}, {Collins}, {Girardin}, {Heldridge}, {Kotten}, {Mancini}, {Murgas}, {Narita}, {Radford}, {Relles}, {Shporer}, {Soares-Furtado}, {Strakhov}, {Ziegler}, {Brice{\~n}o}, {Calkins}, {Clark}, {Collins}, {Esquerdo}, {Fajardo-Acosta}, {Fukui}, {Watkins}, {He}, {Horne}, {Jenkins}, {Mann}, {Naponiello}, {Palle}, {Schwarz}, {Seager}, {Southworth}, {Srdoc}, {Swift}, \& {Winn}}]{Vowell2025}
{Vowell}, N., {Rodriguez}, J.~E., {Latham}, D.~W., {et~al.} 2025, arXiv e-prints, arXiv:2501.09795, \dodoi{10.48550/arXiv.2501.09795}

\bibitem[{{Wilson} {et~al.}(2004){Wilson}, {Henderson}, {Herter}, {Matthews}, {Skrutskie}, {Adams}, {Moon}, {Smith}, {Gautier}, {Ressler}, {Soifer}, {Lin}, {Howard}, {LaMarr}, {Stolberg}, \& {Zink}}]{2004SPIE.5492.1295W}
{Wilson}, J.~C., {Henderson}, C.~P., {Herter}, T.~L., {et~al.} 2004, in \procspie, Vol. 5492, Ground-based Instrumentation for Astronomy, ed. A.~F.~M. {Moorwood} \& M.~{Iye}, 1295--1305, \dodoi{10.1117/12.550925}

\bibitem[{{Wright} {et~al.}(2010){Wright}, {Eisenhardt}, {Mainzer}, {Ressler}, {Cutri}, {Jarrett}, {Kirkpatrick}, {Padgett}, {McMillan}, {Skrutskie}, {Stanford}, {Cohen}, {Walker}, {Mather}, {Leisawitz}, {Gautier}, {McLean}, {Benford}, {Lonsdale}, {Blain}, {Mendez}, {Irace}, {Duval}, {Liu}, {Royer}, {Heinrichsen}, {Howard}, {Shannon}, {Kendall}, {Walsh}, {Larsen}, {Cardon}, {Schick}, {Schwalm}, {Abid}, {Fabinsky}, {Naes}, \& {Tsai}}]{wright2010}
{Wright}, E.~L., {Eisenhardt}, P. R.~M., {Mainzer}, A.~K., {et~al.} 2010, \aj, 140, 1868, \dodoi{10.1088/0004-6256/140/6/1868}

\bibitem[{{Wu} {et~al.}(2025){Wu}, {Hadden}, {Dewberry}, {El-Badry}, \& {Matzner}}]{Wu2025}
{Wu}, Y., {Hadden}, S., {Dewberry}, J., {El-Badry}, K., \& {Matzner}, C.~D. 2025, \apjl, 982, L34, \dodoi{10.3847/2041-8213/adb751}

\bibitem[{{Zhang} {et~al.}(2025){Zhang}, {Carmichael}, {Huber}, {Stassun}, {Fukui}, {Narita}, {Murgas}, {Palle}, {Latham}, {Calkins}, {Seager}, {Winn}, {Vezie}, {Hounsell}, {Osborn}, {Caldwell}, \& {Jenkins}}]{Zhang2025}
{Zhang}, E.~Y., {Carmichael}, T.~W., {Huber}, D., {et~al.} 2025, arXiv e-prints, arXiv:2503.05115, \dodoi{10.48550/arXiv.2503.05115}

\bibitem[{{Zhou} {et~al.}(2019){Zhou}, {Bakos}, {Bayliss}, {Bento}, {Bhatti}, {Brahm}, {Csubry}, {Espinoza}, {Hartman}, {Henning}, {Jord{\'a}n}, {Mancini}, {Penev}, {Rabus}, {Sarkis}, {Suc}, {de Val-Borro}, {Rodriguez}, {Osip}, {Kedziora-Chudczer}, {Bailey}, {Tinney}, {Durkan}, {L{\'a}z{\'a}r}, {Papp}, \& {S{\'a}ri}}]{Zhou2019hatp70}
{Zhou}, G., {Bakos}, G.~{\'A}., {Bayliss}, D., {et~al.} 2019, \aj, 157, 31, \dodoi{10.3847/1538-3881/aaf1bb}

\bibitem[{{Ziegler} {et~al.}(2020){Ziegler}, {Tokovinin}, {Brice{\~n}o}, {Mang}, {Law}, \& {Mann}}]{Ziegler2020AJ....159...19Z}
{Ziegler}, C., {Tokovinin}, A., {Brice{\~n}o}, C., {et~al.} 2020, \aj, 159, 19, \dodoi{10.3847/1538-3881/ab55e9}

\end{thebibliography}
\bibliographystyle{aasjournal}



\end{document}